\documentclass[12pt,preprint,letterpaper,nofootinbib,superscriptaddress]{revtex4}
\pdfoutput=1
\usepackage[utf8]{inputenc}
\usepackage{epsf, array, xcolor}
\usepackage{graphicx}
\usepackage{subfigure}
\usepackage{bbold}
\usepackage{amsmath,amssymb,mathtools}
\usepackage{epigraph}
\usepackage[colorlinks=true,linktocpage=true,linkcolor=blue,citecolor=blue,urlcolor=blue]{hyperref}
\usepackage{textcomp}
\usepackage{url}

\usepackage{dcolumn}

\begin{document}
\pagestyle{plain}


\title{Uncovering the High Scale Higgs Singlet Model }

\author{Sally Dawson}
\affiliation{Department of Physics, Brookhaven National Laboratory, Upton, N.Y., 11973,  U.S.A.}
\author{Pier Paolo Giardino}
\affiliation{\mbox{Instituto Galego de F\'isica de Altas Enerx\'ias, Universidade de Santiago de Compostela,}\\ \mbox{15782 Santiago de Compostela, Galicia, Spain}}
\author{Samuel Homiller}
\affiliation{Department of Physics, Harvard University, Cambridge, MA, 02138, U.S.A.\vspace{0.25cm}}

\date{\today \vspace{0.5cm}}

\begin{abstract}
The scalar singlet model extends the Standard Model with the addition of a new gauge singlet scalar.  
We re-examine the limits on the new scalar from oblique parameter fits and from a global fit to precision electroweak observables and present analytic expressions for our results. 
For the case when the new scalar is much heavier than the weak scale, we map the model onto the dimension-six Standard Model effective field theory (SMEFT) and review the allowed parameter space from unitarity considerations and from the requirement that the electroweak minimum be stable.
A global fit to precision electroweak data, along with LHC observables, is used to constrain the parameters of the high scale singlet model and we determine the numerical effects of performing the matching at both tree level and 1-loop.
\end{abstract}

\maketitle

\section{Introduction}

The Higgs singlet model~\cite{Bowen:2007ia, O'Connell:2006wi, Dawson:2015haa, Muhlleitner:2020wwk, Robens:2015gla, Costa:2015llh, Chen:2014ask} has been extensively studied as a simple extension of the Standard Model (SM) containing only one new particle. 
Depending on the potential parameters, the model can lead to a first order electroweak phase transition~\cite{Huber:2006wf, Profumo:2007wc, Espinosa:2011ax, Barger:2011vm, Profumo:2014opa, Curtin:2014jma, Kotwal:2016tex, Huang:2017jws, Chen:2017qcz, Kurup:2017dzf, Li:2019tfd}, making it highly motivated in addressing the problem of baryogenesis.
It can also arise as the limiting case of many interesting models addressing the hierarchy problem~\cite{Craig:2013xia, Curtin:2015bka} or even dark matter~\cite{Silveira:1985rk, McDonald:1993ex, Burgess:2000yq, Menon:2004wv, He:2008qm, Gonderinger:2009jp, Mambrini:2011ik}.
When the mass of the new scalar becomes much larger than the weak scale, the theory can be mapped onto an effective field theory. 
The utility and simplicity of the model thus makes it an ideal candidate for exploring the limits of an effective field theory framework in reproducing the features of the underlying UV models~\cite{Dawson:2020oco, Brehmer:2015rna,  Henning:2014gca, Henning:2014wua, Gorbahn:2015gxa, Ellis:2018gqa,Ellis:2020unq,Anisha:2020ggj}.

In the full UV complete singlet model, restrictions on the parameters can be found from fits to precision electroweak observables as well as LHC data.
These limits can then be compared with limits found in the context of a low energy effective field theory.  We consider an effective field theory in which the SM Higgs doublet is constrained to be
an $SU(2)$ doublet, the Standard Model effective field theory (SMEFT).
At tree level, the singlet model generates only two SMEFT coefficients when matched at the UV scale~\cite{Henning:2014wua, Egana-Ugrinovic:2015vgy}.  The aim of this work is to examine to what extent the extraction of SMEFT coefficients from global fits at the weak scale gives information on the parameters of the UV complete singlet model~\cite{Ellis:2018gqa, Ellis:2020unq, Dawson:2020oco, Bakshi:2020eyg, Kribs:2017znd, Falkowski:2015iwa, Gorbahn:2015gxa}.
The focus is on understanding the numerical importance of various choices made when performing the low energy fits and to this end, we implement both tree and 1-loop matching~\cite{Jiang:2018pbd, Haisch:2020ahr, Cohen:2020fcu} at the UV scale. 
We find that the effects of the 1-loop matching are typically rather small. 
Effects of ${\cal{O}}(10\%)$ can be obtained only for rather large values of certain dimensionless parameters in the Lagrangian.  
 
Section~\ref{sec:bas} contains a recap of the model and restrictions on the model parameters from unitarity and the minimization of the potential. 
Analytic results for electroweak precision observables in the singlet model are found in  Section~\ref{sec:glob} along with a comparison between a global fit to electroweak precision observables (EWPOs) and a fit to the oblique parameters, and restrictions from unitarity and the minimization of the potential are in Section~\ref{sec:lims}.
The SMEFT matching with the singlet model at both tree and loop level is studied  in Section~\ref{sec:fits} and a global fit to electroweak precision observables, Higgs, and di-boson data is presented.
Section~\ref{sec:concl} has some conclusions.

\section{Basics}
\label{sec:bas}

The singlet model we consider contains the SM Higgs doublet, $\Phi$, and a scalar gauge singlet, $S$. The most general scalar potential is,
\begin{align}
V(\Phi, S) = & -\mu_H^2\Phi^\dagger\Phi+\lambda_H(\Phi^\dagger\Phi)^2 \nonumber \\[4pt]
&  + \frac{m_\xi}{2}\Phi^\dagger\Phi S+\frac{\kappa}{2}\Phi^\dagger\Phi S^2 \nonumber \\
&  + t_S S+\frac{M^2}{2}S^2+\frac{m_\zeta}{3}S^3 +\frac{\lambda_S}{4}S^4\, .
\end{align}
The parameters can be redefined such that $\langle {S} \rangle\equiv x=0$. 
After spontaneous symmetry breaking, the 2 neutral scalars, $\Phi_0$ and ${ S}$, mix  to
form the physical scalars, $h$ and $H$,
\begin{eqnarray}
h&=& \cos\theta \,{\Phi_0}+\sin\theta \,{S}\nonumber \\
H&=& -\sin\theta \, {\Phi_0}+\cos\theta\, {S}\, ,
\end{eqnarray}
with the physical masses, $m_h=125.1\,\textrm{GeV}$ and $M_H$.  
The parameters of the model can be taken as,
\begin{equation}
m_h,\quad M_H,\quad v=246\,\textrm{GeV},\quad \sin\theta,\quad x=0,\quad \kappa,\quad m_\zeta,\quad \lambda_S\, .
\label{eq:pardef}
\end{equation}
The other parameters of the Lagrangian are determined  in the singlet model by:\footnote{We note that for $\kappa v \gg M$, the mass of the new scalar,  $M_H$, comes from electroweak symmetry breaking and in this case the theory cannot be mapped onto the SMEFT~\cite{Buchalla:2016bse, Cohen:2020fcu}. 
Additionally, the kinematic distributions for $hh$ production in this limit are quite different from those where $M_H$ primarily depends on $M$~\cite{Dawson:2015oha}.}
\begin{eqnarray}
m_\xi &=& \frac{m_h^2-M_H^2}{v}\sin 2\theta\, , \nonumber \\[4pt]
M^2&=&m_h^2\sin^2\theta+M_H^2\cos^2\theta - \frac{\kappa}{2}v^2\, , \nonumber\\
\lambda_H&=&\frac{m_h^2\cos^2\theta+M_H^2\sin^2\theta}{2 v^2}\, .
\label{eq:lagpar}
\end{eqnarray}
The $Z_2$ symmetric case has $m_\zeta=t_S=m_\xi=0$ and $x\ne 0$.

The couplings of $h$ to SM fermions and gauge bosons are suppressed relative to the SM Higgs couplings by a factor of $\cos\theta$, while the $H$ couplings are suppressed by $\sin\theta$. 
We can thus immediately find a trivial limit on $\cos\theta$ from Higgs production to SM particles $X$, (assuming no decays to invisible particles),\footnote{If $2M_H < m_h$ then the decay $h\rightarrow HH$ is allowed, altering the limit on $\cos~\theta$.} 
\begin{eqnarray}
\cos^2\theta & = &\mu \equiv
\frac{\sigma\cdot \textrm{BR}}{(\sigma\cdot \textrm{BR})_{\textrm{SM}}}
\end{eqnarray}
Naively combining the combined  ATLAS results with $80\,\textrm{fb}^{-1}$~\cite{Aad:2019mbh} and the CMS combined limits with $139\,\textrm{fb}^{-1}$~\cite{CMS:2020gsy},
\begin{equation}
\mu[\textrm{ATLAS}]=1.11^{+.09}_{-.08},
\qquad
\mu[\textrm{CMS}]=1.02^{+.07}_{-.06}
\end{equation}
we find at $95\%$ C.L., 
\begin{equation}
|\sin\theta\, | < 0.2\quad \text{for} \quad m_h < 2 M_H\, .
\label{eq:hbound}
\end{equation}
For $m_h>2M_H$, the naive limit of Eq.\,\eqref{eq:hbound} does not apply because the $h$ decays to $HH$ must be included and this branching ratio is sensitive to the other parameters of the scalar potential. 
Limits on the singlet model from resonant double Higgs production are beginning to be competitive with those from single Higgs production for $M_H\lesssim 700\,\textrm{GeV}$~\cite{Aad:2019uzh}, although our primary focus here will be on $M_H\sim (1-2)\,\textrm{TeV}$.

\section{Restrictions on Model Parameters}
\label{sec:glob}

The parameters  of the singlet model can be  limited by a fit to the $Z$- and $W$-pole observables (we term this the EWPO fit):
\begin{eqnarray}
&&M_W,~~ \Gamma_W,~~ \Gamma_Z,~~  \sigma_h,~~  R_b,~~ R_c,~~ R_l,~~  A_{FB,b},~~ A_{FB,c},~~ A_{FB,l},~~  A_b,~~ A_c,~~ A_l\, .
\label{eq:zpole}
\end{eqnarray} 
The SM results for these observables are well known~\cite{Hollik:1988ii,Freitas:2014hra}.
In a previous study, Ref.~\cite{Dawson:2019clf}, we computed the limits on the coefficients of an effective field theory that result from a fit to the observables
of Eq.\,\eqref{eq:zpole} computed to NLO in both QCD and electroweak interactions, and
we apply an identical calculational framework here.
The observables and SM theory numbers used in the current study  can be found in in Table~III of Ref.~\cite{Dawson:2019clf}. 
We take as our input parameters:
$G_\mu=1.1663787(6)\times 10^{-5}\,\textrm{GeV}^{-2}$,
$M_Z=91.1876\pm .0021\,\textrm{GeV}$, 
$1/\alpha = {137.035999139(31)}$, 
$\Delta\alpha_{\rm had}^{(5)} = 0.02764\pm 0.00009$, 
$\alpha_s(M_Z)=0.1181\pm 0.0011$, ~
$m_h=125.10 \pm 0.14\,\textrm{GeV}$,~
$m_b= 4.18\,\textrm{GeV}$ and 
$M_t=172.9\pm0.5\,\textrm{GeV}$ .

The one loop relation between the Fermi constant $G_\mu$ and the vacuum expectation value $v$ is, as usual, 
\begin{equation}
G_\mu=\frac{1}{\sqrt{2}v^2} (1+\Delta r)
\end{equation}
where,
\[
\Delta r=\Delta r_{\textrm{SM}}+\Delta r_{\textrm{singlet}} . 
\]
In computing $\Delta r$, we use ${\hat {M}}_W^2 \equiv (M_Z^2 / 2) \bigg( 1 + \sqrt{1-\frac{\sqrt{8}\pi\alpha}{G_\mu M_Z^2}}\bigg)$
calculated from our inputs.
For simplicity, we define $h\equiv m_h^2$, $H\equiv M_H^2$, $z\equiv M_Z^2$ and $w\equiv {\hat{M}}_W^2$ and obtain
the simple form,\footnote{The function $A_{0}$ is defined as
 \[
 A_0(m^2)=\int \frac{d^dk}{(2\pi)^2}\frac{1}{k^2-m^2},
 \]
where we calculate in $d=4-2\epsilon$ dimensions.
$B_{0}$ is the Passarino- Veltman $2$-point function,
\[
B_0(m_1^2,m_2^2,p^2)=\int \frac{d^d k}{(2\pi)^n} \frac{1}{[k^2-m_1^2][(k+p)^2-m_2^2]}\, .
\]
The Passarino-Veltman functions are evaluated using \texttt{QCDLOOPS}~\cite{Carrazza:2016gav}.
}
\begin{align}
\Delta r_{\textrm{singlet}} & = \Delta r_{\textrm{singlet}}(h,H)-\Delta r_{\textrm{singlet}}(H,h)\nonumber \\[6pt]
\Delta r_{\textrm{singlet}}(h,H) & = 
{\sqrt{2}\sin^2\theta G_\mu\over 16\pi^2}\biggl\{ 
-{h\over 2  }+ {3wA_{0}(h)\over (h-w)} 
     + {3w hA_{0}(w)\over (H - w)(h-w )}\biggr\}\, .
\end{align}
 
We find the one-loop prediction for $M_W$  in the singlet model,
\begin{eqnarray}
 M_W&=& 
M_{W}^{\textrm{SM}}+F_W(h,H)-F_W(H,h)
\nonumber \\[6pt]
F_W(h,H)&=&
{\alpha\sin^2\theta\over 8\pi M_W}\biggl\{
{hz\over 24(2w - z)} + 
{A_{0}(h)\over 12(h - w)w(2w - z)}
\biggl( hw(w - 4z) + 12w^2z + h^2(z-w)\biggr)
\nonumber \\ &&
 + { h zA_{0}(w)\over 12(h - w)w(H -w)(z-w)(2w - z)}
  \biggl( (2w-z)\biggl[hH-w(h+H)\biggr] 
+w^2(8z-7w)
  \biggr)
  \nonumber \\ &&
  + 
  { h wA_{0}(z)\over 12(z-w)(z-2w)} 
  -
 {z B_{0}(w, h, w)\over 12w(w - z)}
   \biggl( h^2 - 4hw + 12w^2\biggr)
    \nonumber \\ &&
     + 
  {wB_{0}(z, h, z)\over 12(z-2w)(z-w)}
  \biggl(h^2 - 4hz + 12z^2\biggr)
  \biggr\}\, .
\end{eqnarray}
This is in agreement with Ref.~\cite{Lopez-Val:2014jva}.
For a massless $b$ quark, the total $W$ decay width is,
\begin{align}
 \Gamma_W = &~~ \Gamma_W^{\textrm{SM}}+G_W(h,H)-G_W(H,h)
 \nonumber \\[6pt]
G_W(h,H) = &  -{3\sqrt{w} G_F^2\sin^2\theta\over 32\pi^3}
\biggl\{
{wh(-hH-8w^2+4wh+4wH)
\over 2(h - 4w)(H - 4w)} 
\nonumber \\ &- 
{A_{0}(h)\over h(h - 4w)(h - w)}
   \biggl(h^4 - 7h^3w + 21h^2w^2 - 32hw^3 + 8w^4\biggr)
   \nonumber \\ & 
   +{h A_{0}(w)\over 
    (h - 4w)(h - w)(H - w)(H - 4w)}
 \biggl(
 4w^2(H^2+h^2)-5whH(H+h)
 \nonumber \\ & 
 +h^2H^2
 -4w^3(H+h)+18w^2hH-36w^4
 \biggr)    \nonumber \\ &
     + 
     {B_{0}(w, h, w)\over (h - 4w)}
     \biggl(h^3 - 7h^2w + 20hw^2 - 28w^3\biggr)
   \biggr\}\, .
\end{align}
Analytic expressions for the remaining observables of Eq.\,\eqref{eq:zpole}  are given in the supplemental
material attached to this note.

The finite $b$ mass contribution to $Z$ decays to bottom pairs is sensitive to the Higgs-$b$ Yukawa
coupling and generates non-oblique contributions.   We compute $R_l,R_b,R_c$ and $\Gamma_Z$ for 
 $m_b\ne 0$  and find that the numerical effect is less than $\sim 2\%$ for $M_H>20\,\textrm{GeV}$, rising to $\sim 5\%$ for $M_H \sim 10\,\textrm{GeV}$,
justifying the neglect of $b$ mass effects in our fits.

We perform a fit, including correlations, to the observables of Eq.\,\eqref{eq:zpole} to determine the maximum allowed value 
of $\sin\theta$ for a given value of $M_H$ including all one-loop contributions.
It is of interest  to compare the complete EWPO fit with the results using the oblique parameters only.  
Using the results of \cite{Dawson:2009yx,Englert:2020gcp}, we find that the differences between the Peskin-Takeuchi \cite{Peskin:1991sw} variables in the Higgs singlet model and the SM take the form,
\begin{align}
\Delta S & = \frac{\sin^2\theta}{12\pi}(\mathcal{G}(H,z)-\mathcal{G}(h,z))\\[4pt]
\Delta T & = \frac{3\sin^2\theta}{16\pi s_W^2 c_W^2}(\mathcal{K}(H)-\mathcal{K}(h))\\[4pt]
\Delta S+\Delta U & = \frac{\sin^2\theta}{12\pi c_W^2}(\mathcal{G}(H,w)-\mathcal{G}(h,w)),
\end{align}
where $s_W^2=1-c_W^2=1-w/z$ is  $\sin^2 \theta_W$ of the electroweak mixing angle and we define,
\begin{align}
\mathcal{K}(h) & = h(\frac{(z-w)}{(h-w)(h-z)}A_0(h)+\frac{A_0(w)}{(h-w)}-\frac{A_0(z)}{(h-z)})\\[4pt]
\mathcal{G}(h,z) & = \frac{h}{2}+\mathcal{F}(h,z)(A_0(h)-A_0(z)-(h-z)B_0(z,h,z))\\[4pt]
\mathcal{F}(h,z) & = \frac{h^2-4 h z +12z^2}{z(h-z)}\, .
\end{align}

We fit to the values in~\cite{Zyla:2020zbs},
\begin{align}
\Delta S & = -0.01 \pm 0.10 \nonumber \\
\Delta T & = 0.02 \pm 0.12 \nonumber \\
\Delta U & = 0.02\pm 0.11\,
\end{align}
with the correlation matrix,
\begin{equation}
\rho=\left(
\begin{array}{c c c}
1.&0.92&-0.80\\
0.92 & 1.&-0.93\\
-0.80 & -0.93&1.
\end{array}\right)\, .
\end{equation}

\begin{figure}[]
\includegraphics[width=.48\textwidth]{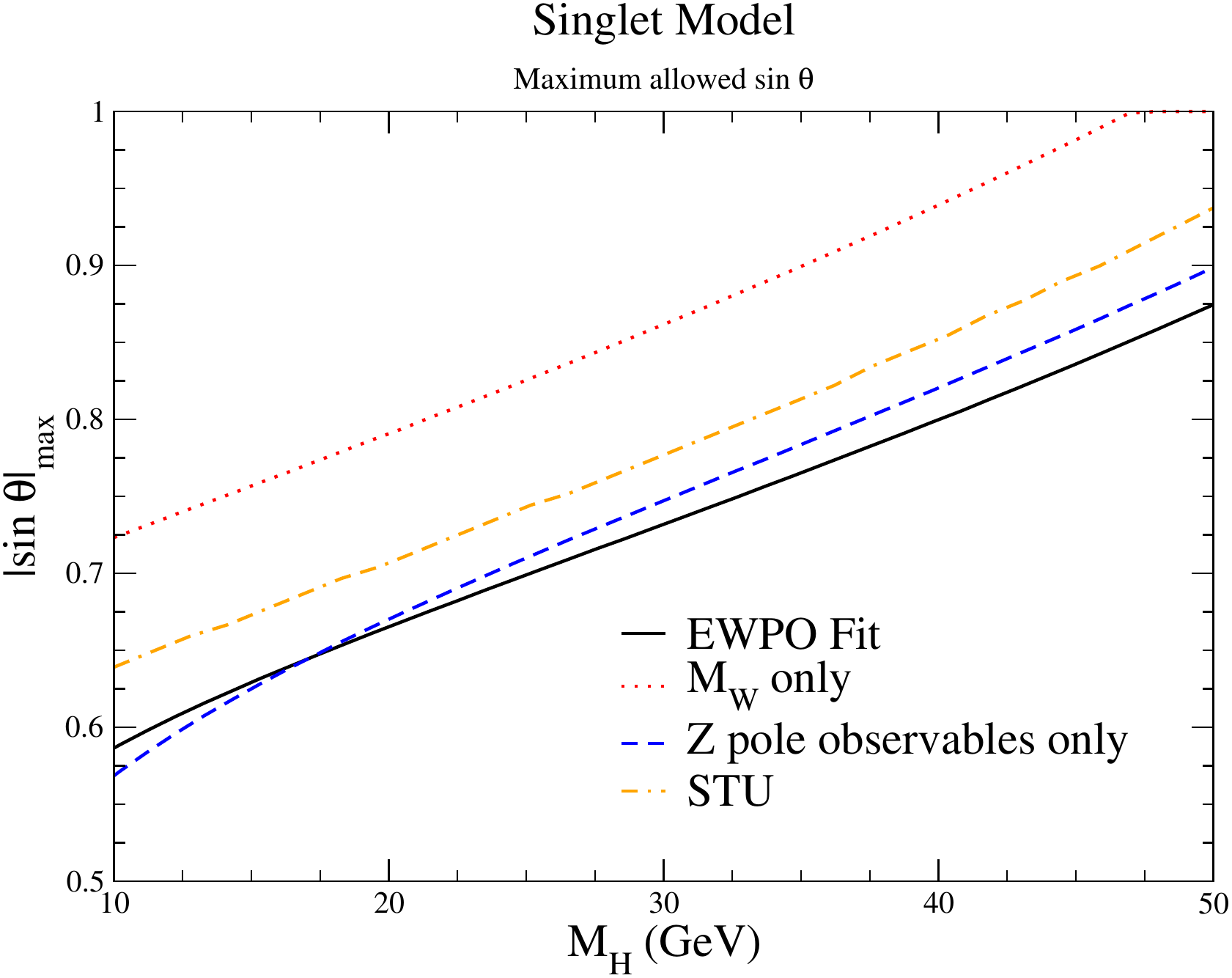}
~
\includegraphics[width=.48\textwidth]{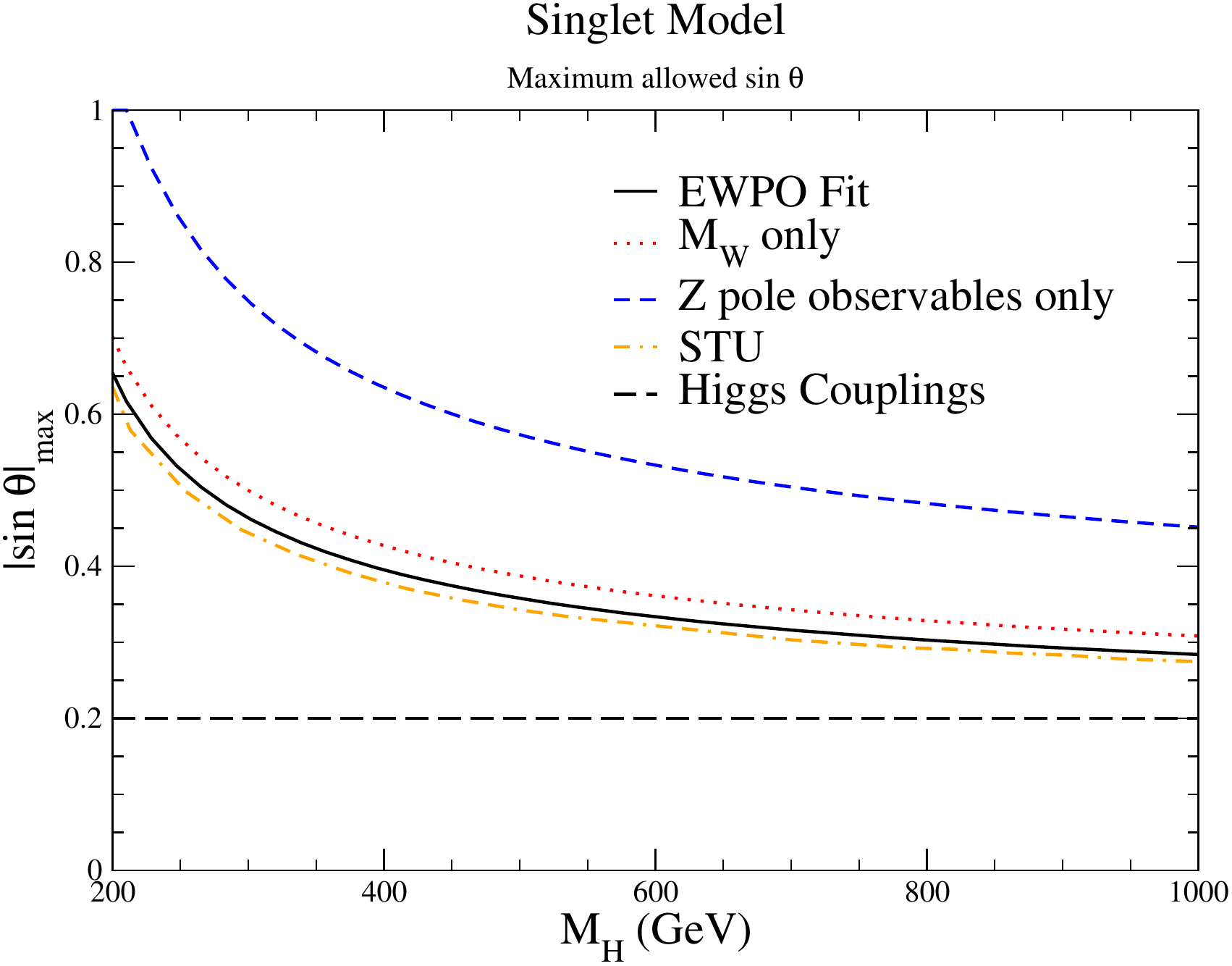}
\caption{ Maximum allowed  $\sin\theta$ in the Higgs singlet model at $95\%$ confidence level based on fits to electroweak precision observables (EWPO), the $W$ mass measurement only, the $Z$ pole observables only, and to the $STU$ parameters as described in the text. A naive limit from the Higgs coupling measurements is shown on the RHS for comparison. }
\label{fig:global}
\end{figure}

In Fig.~\ref{fig:global} we report the results corresponding to different sets of observables:
\begin{itemize} \setlength\itemsep{0.125em}
\item Only $M_W$
\item The $Z$ pole observables alone
\item Oblique parameters only
\item EWPOs given in Eq.\,\eqref{eq:zpole}.
\end{itemize}

The results for the fit to $M_W$ alone are in agreement with those of Ref.~\cite{Ilnicka:2018def}  and are a good approximation to the complete EWPO fit. 
The EWPO fit limits are in agreement with Ref.~\cite{Falkowski:2015iwa} after adjusting for the different input parameters. 
It is interesting that  the current limits from Higgs couplings give better bounds for all $M_H \lesssim 1\,\textrm{TeV}$ as given in Eq.\,\eqref{eq:hbound}.
The limits obtained from oblique parameters are  in approximate agreement with those from the EWPO fit for a heavier
second Higgs boson, although slightly different data sets and approximations were used in the numbers we fit to.\footnote{
We find rough agreement with Refs.~\cite{Profumo:2014opa,Falkowski:2015iwa} 
(the differences can be explained by the different numerical values of the input parameters)
and disagree with the oblique parameter limits of Fig.~1 of \cite{Chalons:2016lyk}. 
We note that the curve labelled ``Exact Singlet'' on the RHS of Fig. 1 of Ref.~\cite{Dawson:2020oco} is the $STU$ result and has used a  slightly different fit to the oblique parameters~\cite{deBlas:2017wmn} from the 
PDG~\cite{Zyla:2020zbs} results used here.
The curve labelled Higgs in that plot is the prediction from fitting Higgs data within the context a SMEFT fit and thus differs from the SM Higgs coupling fit shown in Fig.~\ref{fig:global}.}
For the case where the second Higgs is light, $M_H< m_h$, the limits obtained from the oblique parameters are not a good approximation of the complete EWPO fit.

\section{Theoretical Constraints}
\label{sec:lims}

In Section \ref{sec:fits}, we will match the singlet model with a very heavy $H$ to the SMEFT. Before we do so, we consider the theoretical restrictions on the singlet model parameters 
that are relevant for the matching. 

\subsection{Vacuum Structure of the Potential}

The first set of theoretical constraints on the singlet model come from requiring a suitable vacuum structure of the potential~\cite{Espinosa:2011ax, Chen:2014ask,Robens:2015gla,Robens:2016xkb}. Demanding that the potential is stable at large field values leads to the requirement $\lambda_H, \lambda_S > 0$, and $\kappa \geq -2 \sqrt{\lambda_H \lambda_S}$~\cite{Chen:2014ask},
where $\lambda_H $ is determined by Eq.\,\eqref{eq:lagpar}.
Additional bounds result from requiring that the electroweak minimum be the global minimum of the potential. 
Following \cite{Chen:2014ask}, we compute these bounds by finding all the extrema of the potential expanded around the electroweak vev as a function of $(v, x)$, and then checking whether or not the value of the potential at $(v=246\,\textrm{GeV}, x=0)$ is the global minimum.

\begin{figure}
\includegraphics[width=0.55\linewidth]{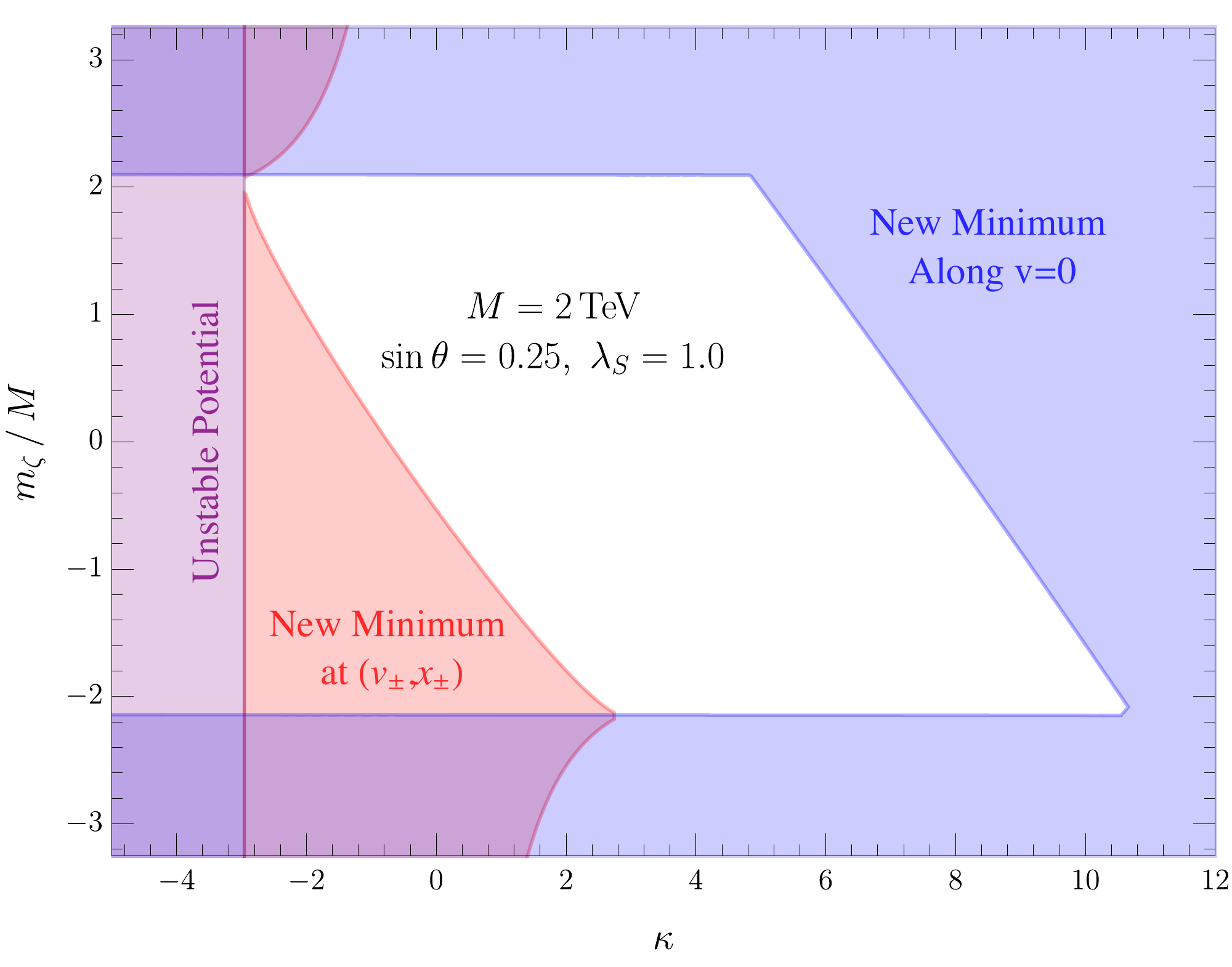}
\caption{Demonstration of the bounds from the appearance of other global minima in the $\kappa$ vs. $m_{\zeta} / M$ plane for $M = 2\,\textrm{TeV}$, $\sin\theta = 0.25$, and $\lambda_S = 1.0$ in the singlet model.}
\label{fig:vacuum_plot_example}
\end{figure}

\begin{figure}
\includegraphics[width=0.45\linewidth]{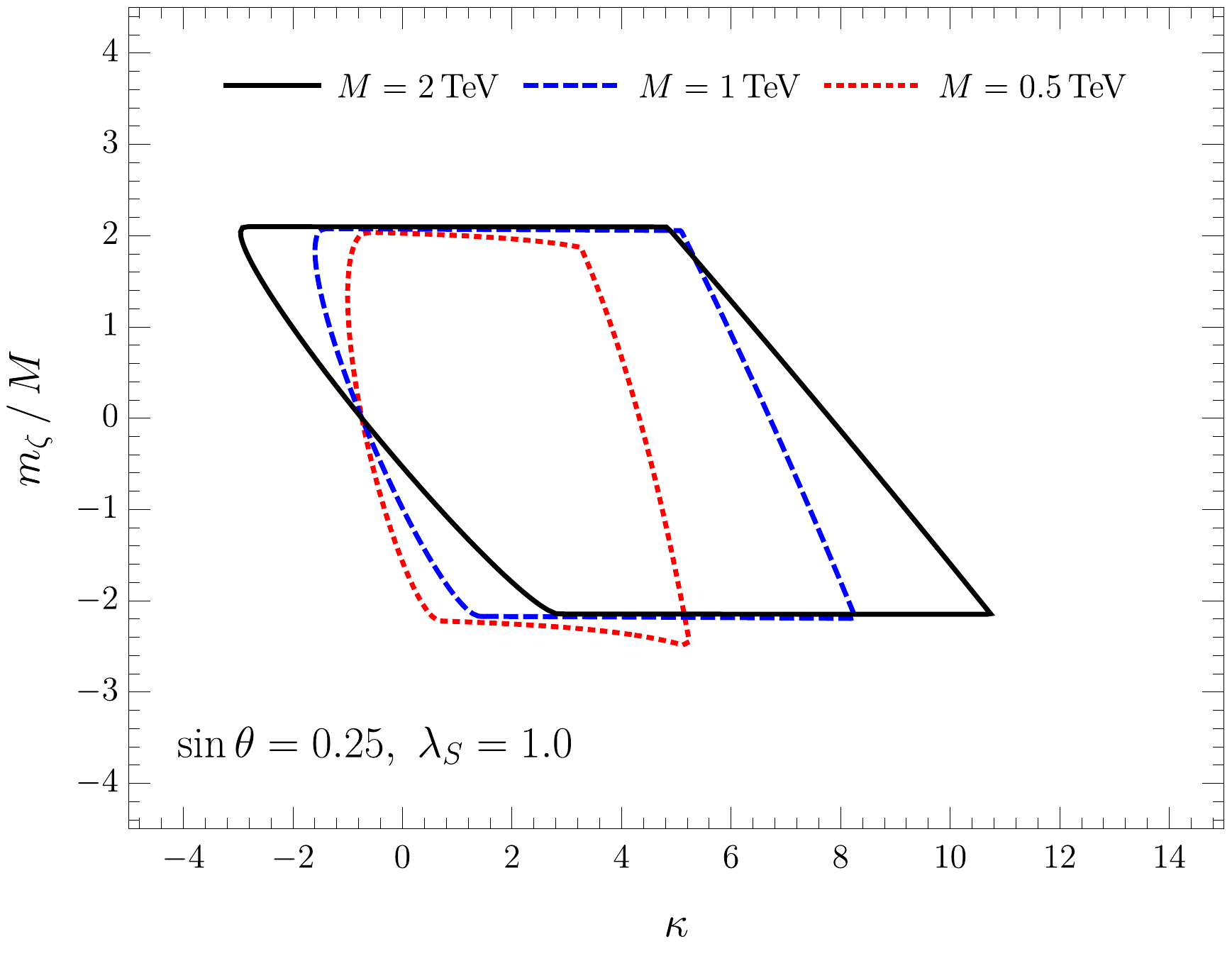}
\quad
\includegraphics[width=0.45\linewidth]{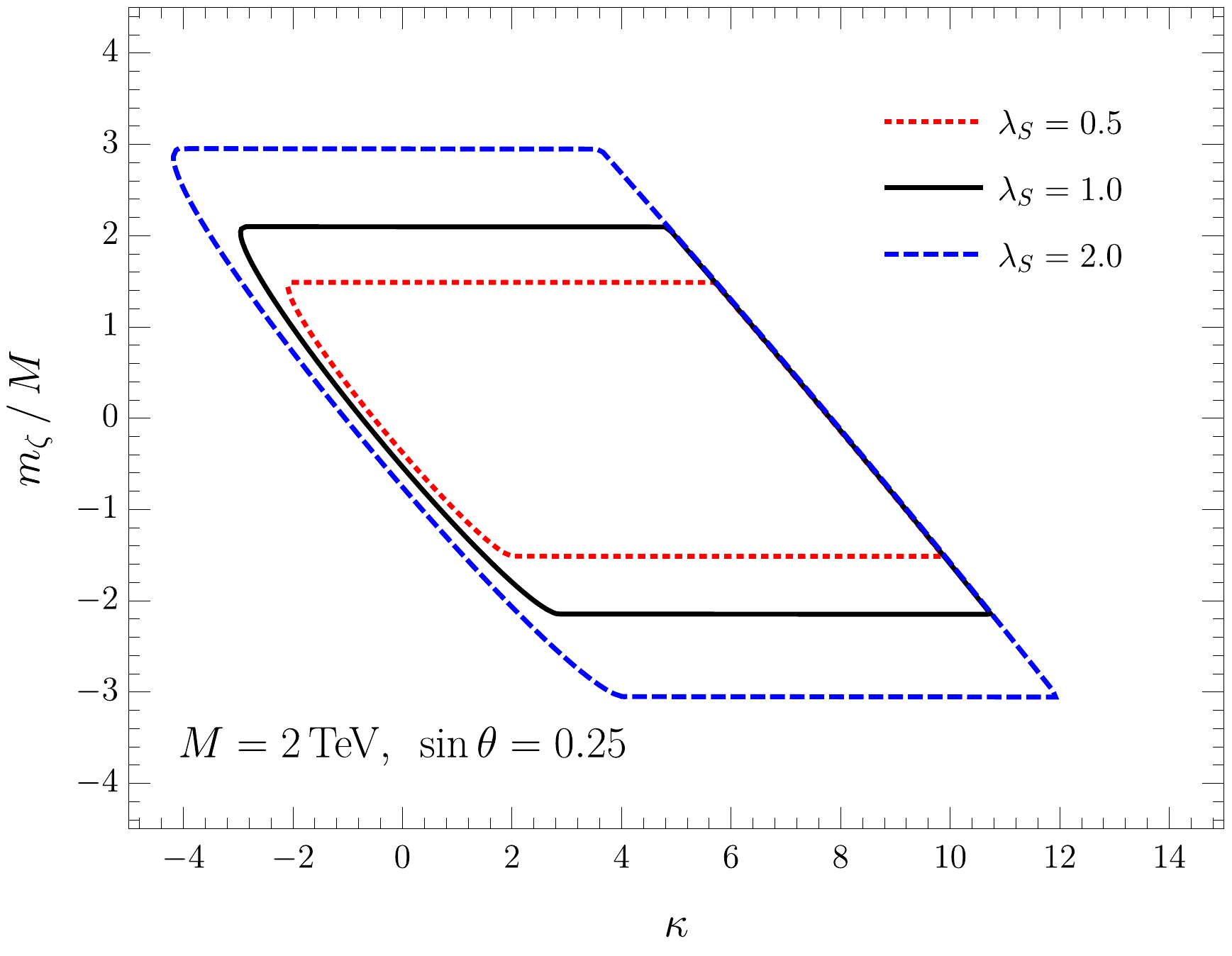}\\
\vskip 0.5cm
\includegraphics[width=0.45\linewidth]{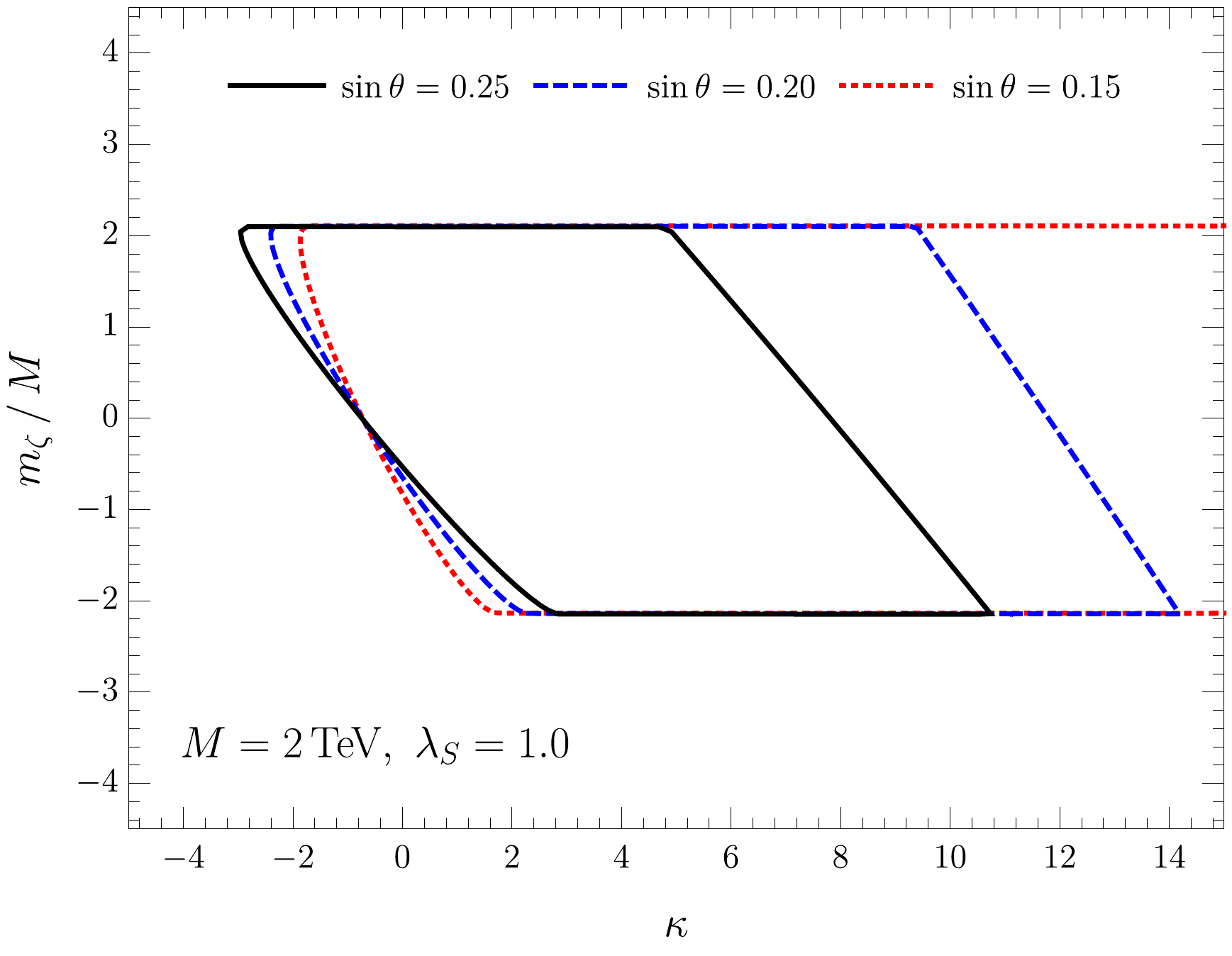}
\caption{Regions in the singlet model where the electroweak minimum is the global minimum of the potential as a function of $\kappa$ and $m_{\zeta} / M$, varying the other physical parameters.}
\label{fig:vacuum_plots}
\end{figure}

The extrema of the potential can be divided into two classes: extrema where $v \neq 0$, and those where $v = 0$. 
In the former case, the new extrema are denoted $(v_{\pm}, x_{\pm})$ as in ref.~\cite{Chen:2014ask}, and tend to bound lower values of $\kappa$.
In the latter case, the extrema are denoted by $(0, x^0_{\pm})$, and these tend to limit both large values of $\kappa$ as well as large values of $m_{\zeta}$. An example of the vacuum structure is shown in Fig.~\ref{fig:vacuum_plot_example}, where we illustrate the regions excluded by the emergence of different global minima as well as the condition from vacuum stability.  
Fig.~\ref{fig:vacuum_plots} illustrates how these bounds change as a function of the physical parameters. In particular, we see that for larger masses, $M$, the bounds on $m_{\zeta} / M$ from the $v = 0$ minima become constant as a function of $\kappa$, depending only on $\lambda_S$.  It is interesting that quite large values of $\kappa$ are allowed in all scenarios.  The upper bound on $m_\zeta/M$ never
exceeds ${\cal{O}}(2-3)$.

\subsection{Unitarity}

The next set of theoretical constraints come from the requirements of tree-level perturbative unitarity~\cite{Lee:1977yc, Lee:1977eg,Dawson:2017vgm,Robens:2015gla}.
The simplest constraints come from $hh \to hh$ and $HH \to HH$ scattering, where the spin-$0$ partial waves in the high energy limit are
\begin{align}
a_0(hh \to hh) \Big|_{s \gg m_h^2} & = - \frac{3}{16\pi}\left( \lambda_H \cos^4{\theta}+ \kappa \sin^2{\theta} \cos^2{\theta} 
+ \lambda_S\sin^4{\theta} \right) \\
a_0(HH \to HH) \Big|_{s \gg m_H^2} & = - \frac{3}{16\pi} \left( \lambda_S \cos^4{\theta} + \kappa \sin^2{\theta} \cos^2{\theta}+ \lambda _H
\sin^4{\theta} \right)
\end{align}
For $\sin\theta \ll 1$, requiring $|a_0| < 1/2$ sets the bounds $\lambda_S, \lambda_H \lesssim 8\pi / 3$. This bound on $\lambda_H$ indirectly bounds $M_H$ as a function of $\sin\theta$:
\begin{eqnarray}
M_H^2\, \sin^2\theta &\lesssim &\frac{16\pi}{3}v^2 - m_h^2\, \cos^2\theta \nonumber \\
M_H&\lesssim& 7\,\textrm{TeV}\quad {\rm{for}}\cos\theta=0.99\, .
\end{eqnarray}
The similar bound from $hH \to hH$ scattering only restricts $|\kappa| \lesssim 8\pi$.

\clearpage
\section{One-loop Matching of the Singlet Model to SMEFT}
\label{sec:fits}

\subsection{One-loop Matching}

When the mass of the heavy scalar is much larger than the weak scale and any relevant energy scales, the singlet model can
be modeled by an effective field theory,
\begin{equation}
\mathcal{L}_{\textrm{singlet}} 
~\xrightarrow[M_H \to \infty]{} ~
\mathcal{L}_{\textrm{SM}} + 
\sum_i \frac{C_i(M)}{M^2} \mathcal{O}_i^{(6)} + \dots
\end{equation} 
with coefficients  matched to the singlet model at the high scale, $M$.  We retain only the dimension-6 operators, $\mathcal{O}_i^{(6)}$ ,and use the Warsaw basis~\cite{Buchmuller:1985jz} with the notation of Ref.~\cite{Dedes:2017zog}.

The global fits of Ref.~\cite{Dawson:2020oco} were performed using tree level matching at the scale $M$.\footnote{Ref.~\cite{Brehmer:2015rna} noted that better agreement between the SMEFT and singlet model predictions for $hh$ production are obtained when the matching is performed at the physical
mass, $M_H$. The one-loop matching would then contain terms proportional to $\log(M_H/M)$ that we have omitted.}
It is of interest to implement the one-loop matching for the case of the singlet model and examine the numerical impacts.
The coefficients at the matching scale, $M$, generically take the form,
\begin{equation}
C_i(M)=c_i(M)+ \frac{d_i(M)}{16 \pi^2}
\, ,\end{equation}
where $c_i(M)$ is the tree level result and $d_i(M) / (16\pi^2)$ is the one-loop contribution at the matching scale.
When the renormalization group evolution to the low scale $\mu_{R}$ is included,
\begin{equation}
C_i(\mu_{R})=c_i(M)+{d_i(M)\over 16 \pi^2}+{\gamma_{ij}\over 32\pi^2}c_{ij}(M)\log\biggl({\mu_{R}^2\over M^2}\biggr)\, .
\end{equation}
In the case of the singlet model only two coefficients are generated at tree level~\cite{Egana-Ugrinovic:2015vgy, Henning:2014wua,deBlas:2017xtg,Dawson:2017vgm},
\begin{eqnarray}
c_{H\square}&=&-{m_\xi^2\over 8 M^2}\\
c_H&=& {m_\xi^2 \over 8 M^2}\biggl({m_\xi m_\zeta\over 3M^2}-\kappa\biggr)\, ,
\end{eqnarray}
with all other $c_i(M)=0$.  However, there are many coefficients generated at one-loop
at the matching scale, $M$~\cite{Jiang:2018pbd,Cohen:2020fcu,Haisch:2020ahr}.
The majority of these coefficients are proportional to the tree level coefficient, $c_{H\square}$.
We use the shorthand $C_{Hu}\rightarrow C_{Hu}, C_{Hc}, C_{Ht}$, etc., and take $y_u=y_c=0,~y_t=M_t\sqrt{2}/v$
(similarly we set all other $y_i=0$) and we further assume that $C_{Hq}^{(1)},~C_{Hq}^{(3)},~C_{Hl}^{(1)}$, and $C_{Hl}^{(3)}$ are flavor diagonal and use an analogous shorthand. 
For convenience, we list the results of 
Ref.~\cite{Jiang:2018pbd} in our notation:\footnote{Since $y_t$ is the only non-zero Yukawa that we include, $O_{2y}=y_t^2{\overline t}t{\overline t}t$.}
\begin{eqnarray}
d_{HD}&=&{31 g^{\prime~2}\over 9} c_{H\square}\nonumber \\
d_{HW}&=&-{g^2\over 6}c_{H\square}\nonumber \\
 d_{HB}&=&-{g^{\prime~2}\over 6} c_{H\square}\nonumber \\
d_{HWB}&=&-{g g^\prime \over 3} c_{H\square}\nonumber\\
 d_{Hu}&=&{1\over 108} ( 34g^{\prime~2}-135 y_u^2)c_{H\square}
\nonumber \\
d_{Hd}&=&{1\over 3}   d_{He}={2 \over 3 }   d_{Hl}^{(1)}
=
-{17 g^{\prime~2}\over 108} c_{H\square}\nonumber \\
d_{Hq}^{(1)}&=&{1\over 216} (17g^{\prime~2}+135 y_u^2)c_{H\square}\nonumber\\
d_{Hq}^{(3)}&=& {1\over 72} \biggl[ 17 g^{~2} -45y_u^2\biggr]  c_{H\square}\nonumber \\
d_{Hl}^{(3)}&=& {17 g^2\over 72} c_{H\square}\nonumber \\
 d_{2y}&=& -{1\over 3} c_{H\square}\, .
\end{eqnarray}
The one-loop contribution $d_{tH}$ can be written  in terms of $c_{H\square}$  and $C_H$ and is,
\begin{equation}
d_{tH}= y_t\bigg[ -{1\over 18} (45 y_t^2 -31g^2)c_{H\square}+{3\over 2 } c_H-{29\over 3}\lambda\, c_{H\square}
\bigg]\, ,
\end{equation}
where in the SMEFT the   physical Higgs mass is determined in terms of the  potential parameters  to
${\cal{O}}(v^2/M^2)$ by~\cite{Dedes:2017zog},
\begin{eqnarray}
{m_h^2\over 2v^2}& =&\lambda_H\biggl( 1 +{2v^2\over M^2}  c_{H\square}\biggr) -{3 \over 2}{v^2\over M^2} c_H 
\label{eq:gg}
\end{eqnarray}
and we define,
\begin{eqnarray}
\lambda&\equiv & {2 \lambda_H} \biggl( 1 +{2v^2\over M^2}  c_{H\square}\biggr)\nonumber \\
\lambda & = &\frac{m_h^2}{v^2} + 3 \frac{v^2}{M^2} c_H + \mathcal{O}\bigg( \frac{v^4}{M^4}\bigg)\, ,
\label{eq:lll}
\end{eqnarray}
where we note that Ref.~\cite{Jiang:2018pbd} absorbs the factor of $c_{H\square}$ into the definition of $\lambda$ used in the matching conditions, along with a relative factor of $2$ in the definition of  the quartic terms in the potential.\footnote{ We drop the $c_{HD}$ term in Eq.\,\eqref{eq:gg} since it doesn't occur in the singlet model.}
Eq.\,\eqref{eq:gg} represents the dimension-6 SMEFT limit of Eq.\,\eqref{eq:lagpar} for the relationship between the parameters of the potential and $m_h$.

Finally, the coefficients generated at tree level also receive one-loop corrections,
\begin{align}
d_{H\square} & =  -\frac{9}{2} \lambda c_{H\square}+{31\over 36}
(3 g^2+g^{\prime~2}) c_{H\square}+{3\over2} c_{H}
+\delta d_{H\square}+\delta d_{H\square}^{shift}\nonumber \\
\nonumber \\ 
d_{H} & = \lambda\biggl[\biggl(
 {62g^2-336\lambda\over 9}\biggr)c_{H\square}
+6c_H
\biggr]
+\delta d_{H}+\delta d_{H}^{shift}\, .
\end{align}
where, 
\begin{align}
\delta d_H = &  -\frac{\kappa^3}{12} 
+ \frac{m_{\xi}}{4M^2} \bigg( 
m_{\xi} \big( 9 \lambda^2 - 12 \kappa \lambda + \frac{11}{2} \kappa^2 - 3 \kappa \lambda_S\big) - \kappa^2 m_{\zeta}\bigg) \nonumber \\
& + \frac{m_{\xi}^2}{6 M^4} \bigg( \frac{m_{\xi}^2}{16} \big( 39 \kappa - 18 \lambda_S - 36 \lambda\big) 
	+ m_{\xi}m_{\zeta} \big( 9 \lambda - \frac{15}{2} \kappa + 3 \lambda_S \big) + 3 \kappa m_{\zeta}^2\bigg) \nonumber \\
& + \frac{m_{\xi}^3}{12 M^6} \bigg(-\frac{1}{8} m_{\xi}^3 - \frac{9}{8} m_{\xi}^2 m_{\zeta} + 3 m_{\xi} m_{\zeta}^2 - 2 m_{\zeta}^3\bigg) \label{eq:ee}\,,\\
\delta d_{H\square} = &  -\frac{\kappa^2}{24} 
+ \frac{ m_{\xi}}{12 M^2} \bigg( \frac{m_{\xi}}{2} \big( 17\kappa - \frac{27}{2}\lambda - 18\lambda_S\big) - 5 \kappa m_{\zeta} \bigg) \nonumber \\
& + \frac{m_{\xi}^2}{24 M^4}\bigg( \frac{13}{8} m_{\xi}^2 - 8 m_{\xi} m_{\zeta} + 11 m_{\zeta}^2 \bigg)\,.
\label{eq:dd}
\end{align}

The terms of Eqs.\,\eqref{eq:ee} and \eqref{eq:dd} can be written in terms of $c_H, c_{H\square}$ along with $m_\zeta, \lambda_S$ and $\kappa$, ($m_\xi$ can be written in terms of these parameters).
The one-loop shift terms from canonically normalizing the Higgs kinetic energy are,
\begin{eqnarray}
\delta d_H^{shift}&=& 3 c_{H\square}c_H\, , \nonumber \\
\delta d_{H\square} ^{shift}&=& 2( c_{H\square})^2\, .
\end{eqnarray}
The one-loop shift terms are ${\cal{O}}({v^4 / M^4})$ and can be neglected, since we consistently work to linear order
in the coefficient functions.  

\begin{figure}[t]
\includegraphics[width=0.5\textwidth]{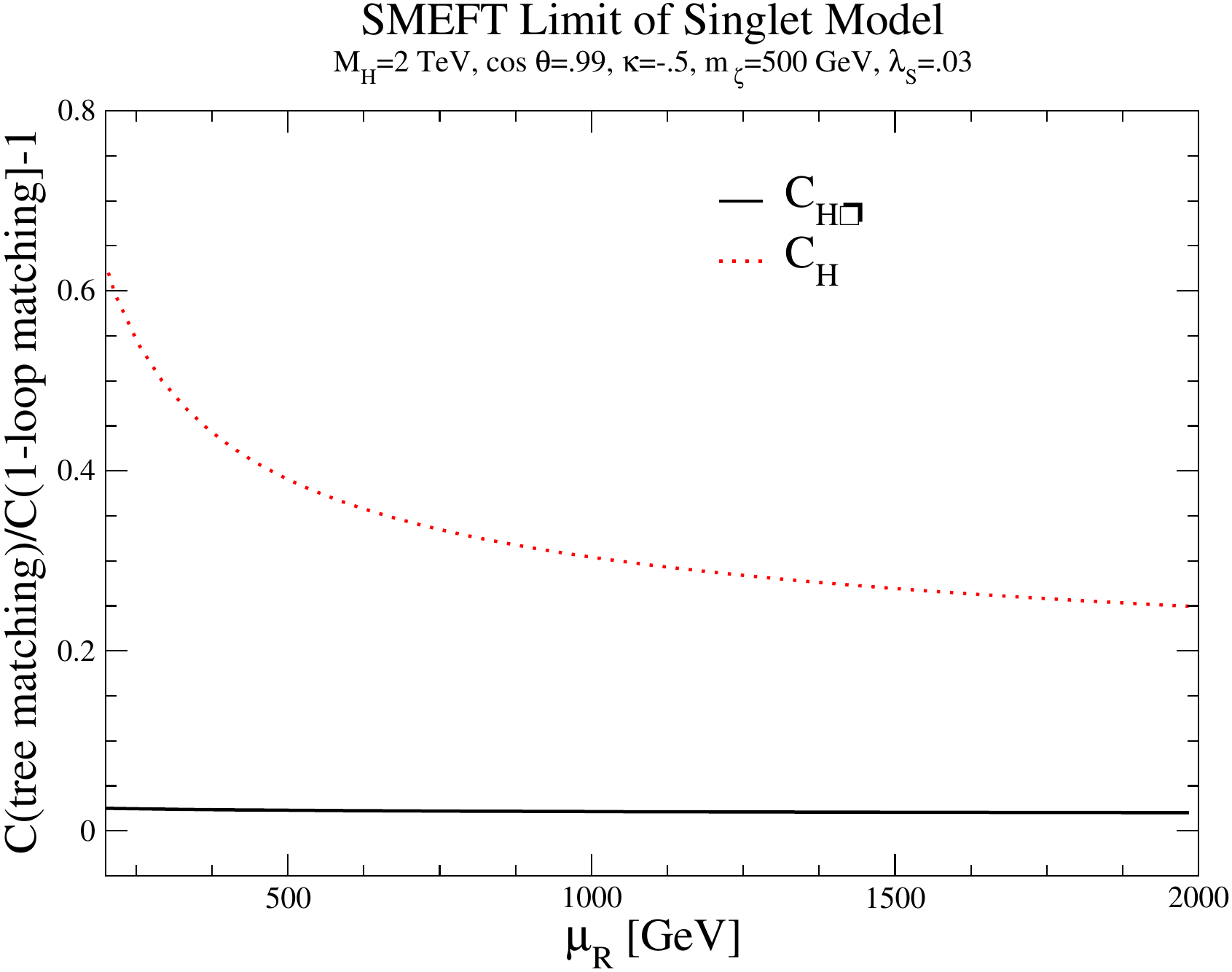}
\caption{Renormalization group evolution of coefficient functions from the matching scale, $M$, to $\mu_R$ when
the matching is done at tree level and at one-loop for coefficients that are generated at tree level. 
The coefficients are evaluated as a function of the running scale, $\mu_R$.}
\label{fig:evolvea} 
\end{figure}

\begin{figure}
\includegraphics[width=0.5\textwidth]{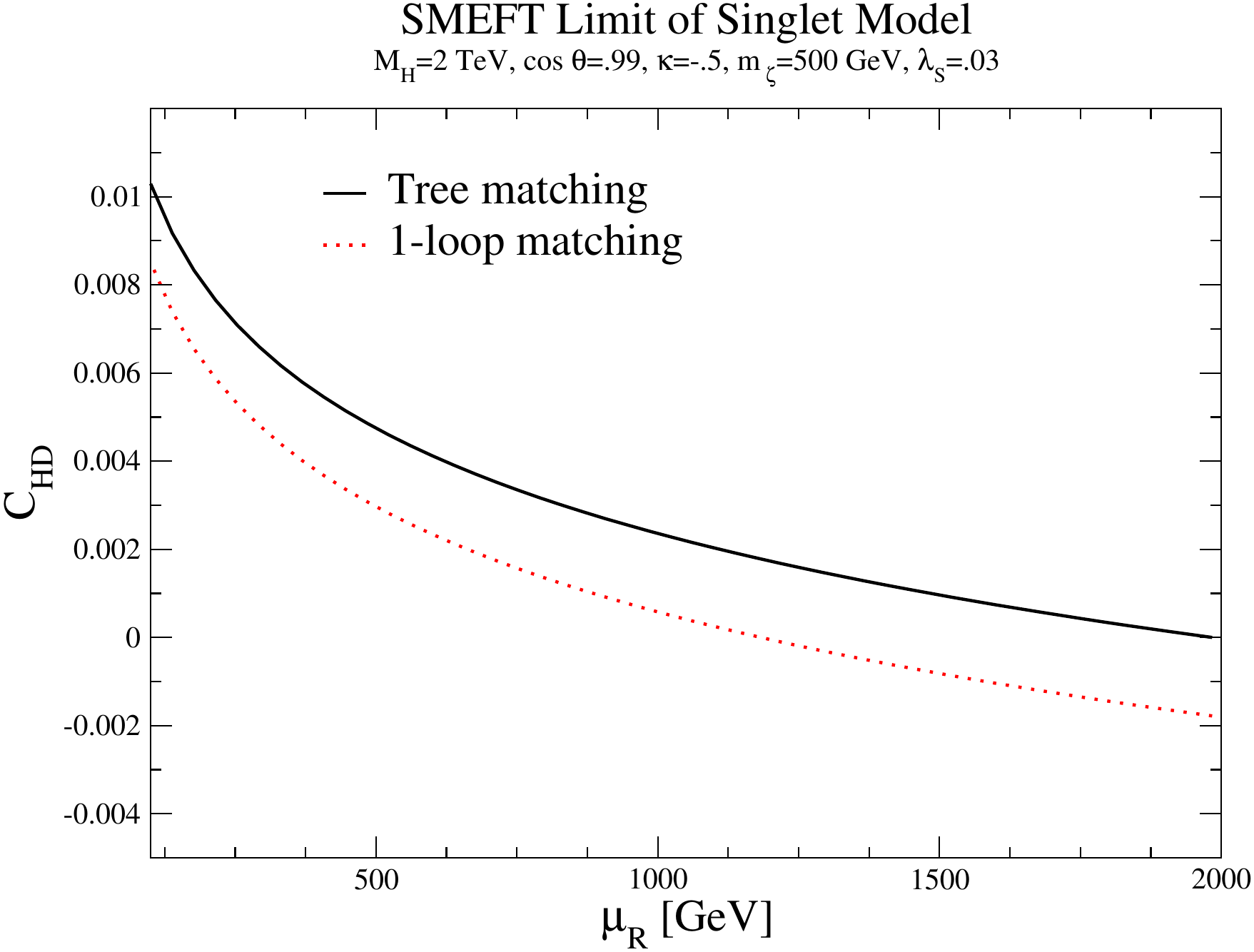}~~
\caption{ Renormalization group evolution of the coefficient function from the matching scale, $M$, to $\mu_R$ for $C_{HD}$, which is generated only by the renormalization group running in the singlet model. 
$C_{HD}$ is evaluated as a function of the running scale, $\mu_R$.}
\label{fig:evolveb} 
\end{figure}

\begin{figure}
\includegraphics[width=0.5\textwidth]{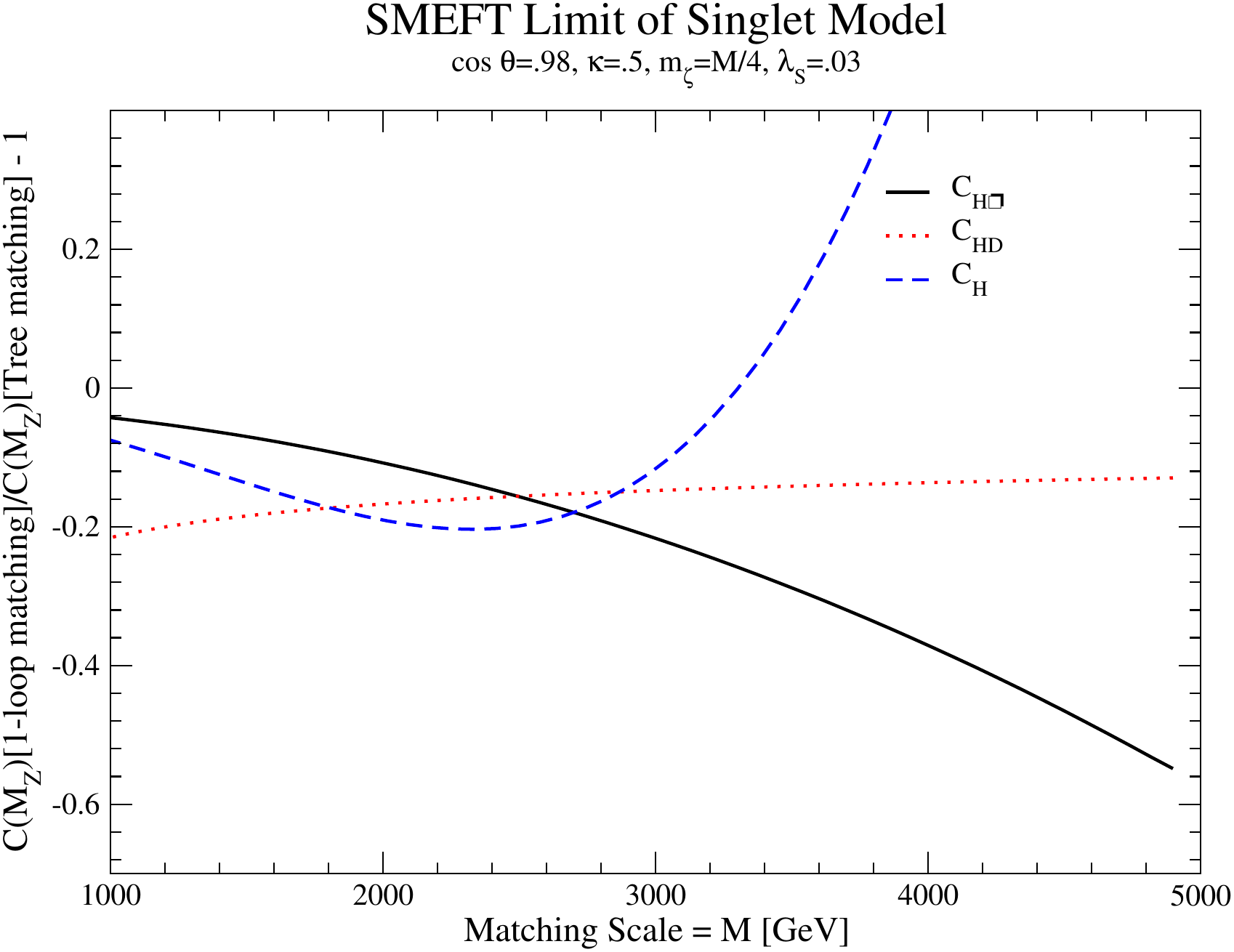}
\caption{ Shift in the coefficient functions at $M_Z$ as a function of the matching scale, $M$,  when
the matching is done at tree level and at one-loop.}
\label{fig:evolve2} 
\end{figure}

After performing the one-loop matching at $M$, the renormalization group is used to evolve the coefficients to $M_Z$,
where the resulting coefficients can be compared with data.\footnote{
A more consistent approach would employ the 2-loop anomalous dimensions, however, these are not available for the SMEFT.
}
The complete set of one-loop anomalous dimension matrices can be found in Refs.~\cite{Jenkins:2013zja, Jenkins:2013wua, Alonso:2013hga}.
The inclusion of the one-loop matching makes  a  relatively minor difference in the evolution of $C_H$ and $C_{H\square}$, as seen in Fig.~\ref{fig:evolvea} where we evolve from $2\,\textrm{TeV}$ 
(note that $M_H$ is related to $M$ by Eq.\,\eqref{eq:lagpar}).
In Fig.~\ref{fig:evolveb}, we show the effect of the one-loop matching on the evolution of $C_{HD}$.
In this case, since $C_{HD}$ is zero at tree level, the contributions from the $1-$loop matching and the renormalization group running are of the same order of magnitude and the effects are more significant. 
In Fig.~\ref{fig:evolve2} we show the relative size of the $1-$loop matching compared to the tree level matching as the matching scale $M$ is increased and the overall effects are between $10-30\%$.
The size of the effects for $C_{H\square}$ and $C_H$ increase dramatically as the matching scale rises over a few TeV.  This is due to the logarithmic running becoming large and in the case of $C_H$, the 1-loop matching terms become of the same order as the tree level terms, implying that the perturbative expansion is no longer valid.  

\subsection{Global Fit}
\label{sec:globs}

Following Ref.~\cite{Dawson:2020oco}, we perform a global fit to the parameters of the non-$Z_2$ symmetric singlet model.  
At the matching scale, $M$, only the tree level coefficients $c_H$ and $c_{H\square}$ are non-zero and other coefficients
are generated at $M_Z$ from the renormalization group running. 
With tree level matching, the results can be expressed in terms of $c_H(M)$ and $c_{H\square}(M)$.  
Using the 1-loop  matching at $M$ described in the previous section, additional coefficients  are generated with a distinctive pattern.
The 1-loop matching introduces a dependence on  three additional parameter combinations beyond those at generated by the tree level matching and  we take as our 5  unknown input parameters, $M_H, \sin\theta, m_\zeta$, $\lambda_S$, and $\kappa$.\footnote{   
The results used to include the effects of $C_H$ require $|C_H| \lesssim (5-6) \big(M / \textrm{TeV}\big)^2$~\cite{Degrassi:2017ucl, Degrassi:2016wml}.} 
The matching scale, $M$, is then calculated using Eq.\,\eqref{eq:lagpar}. 
We match the SMEFT coefficients at $M$ and use the 1-loop renormalization group equations to evolve the SMEFT coefficients to $M_Z$ where we fit to data. 

The included data are identical to that of Ref.~\cite{Dawson:2020oco} and include Higgs coupling strengths from ATLAS~\cite{Aad:2019mbh}, CMS Higgs coupling strengths~\cite{CMS:2020gsy}, $W^+W^-$, $W^\pm Z$, $Wh$ and $Zh$ differential measurements including QCD effects as in~\cite{Baglio:2019uty, Baglio:2020oqu}, and precision electroweak measurements including QCD and electroweak NLO effects from Table~III of Ref.~\cite{Dawson:2019clf}.
We determine the $95\%$ confidence level limits using a $\chi^2$ fit, including the new physics effects at linear order in the SMEFT coefficients.

\begin{figure}[t!]
\centering
\includegraphics[width=0.48\linewidth]{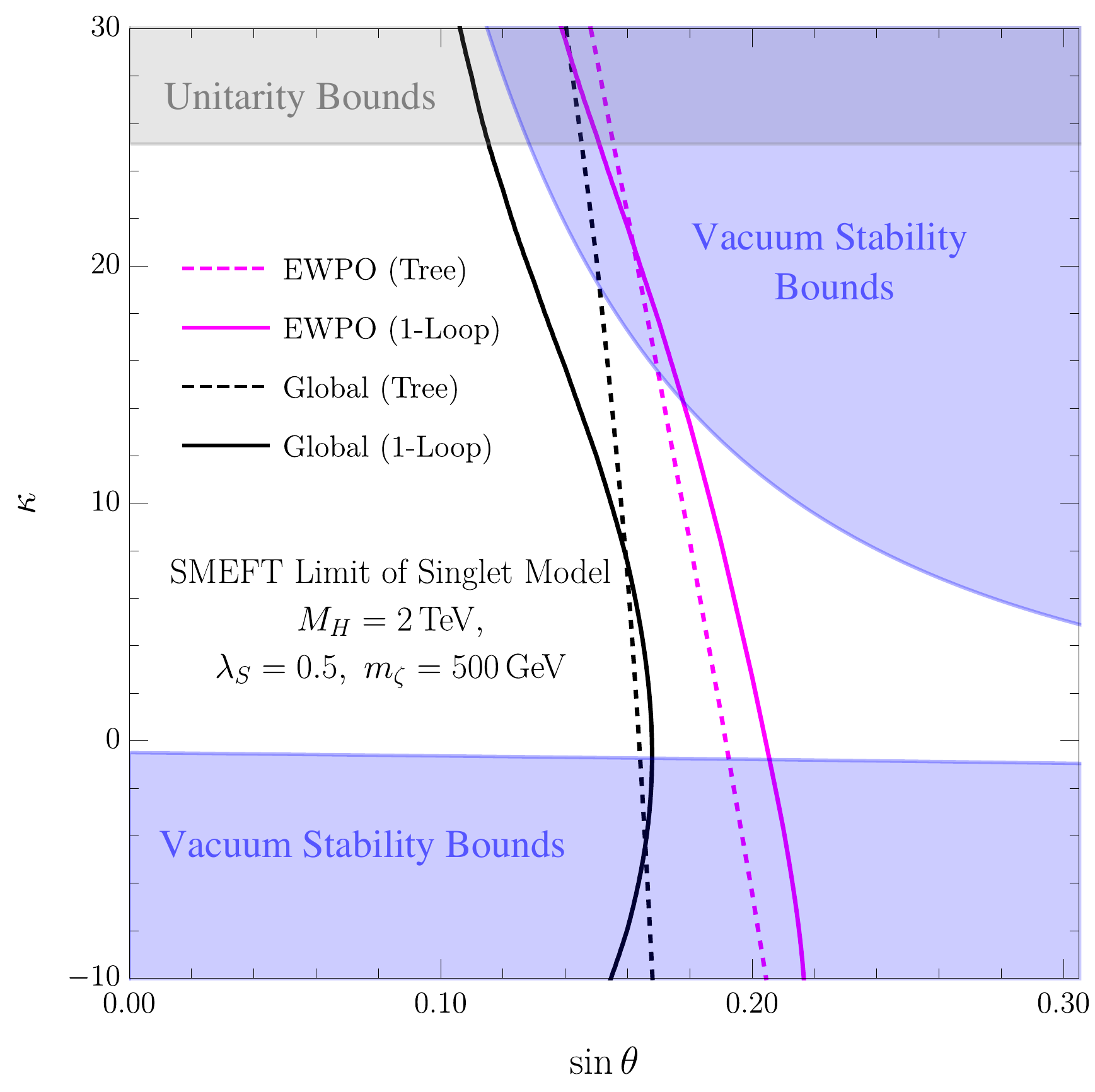}
~
\includegraphics[width=0.48\linewidth]{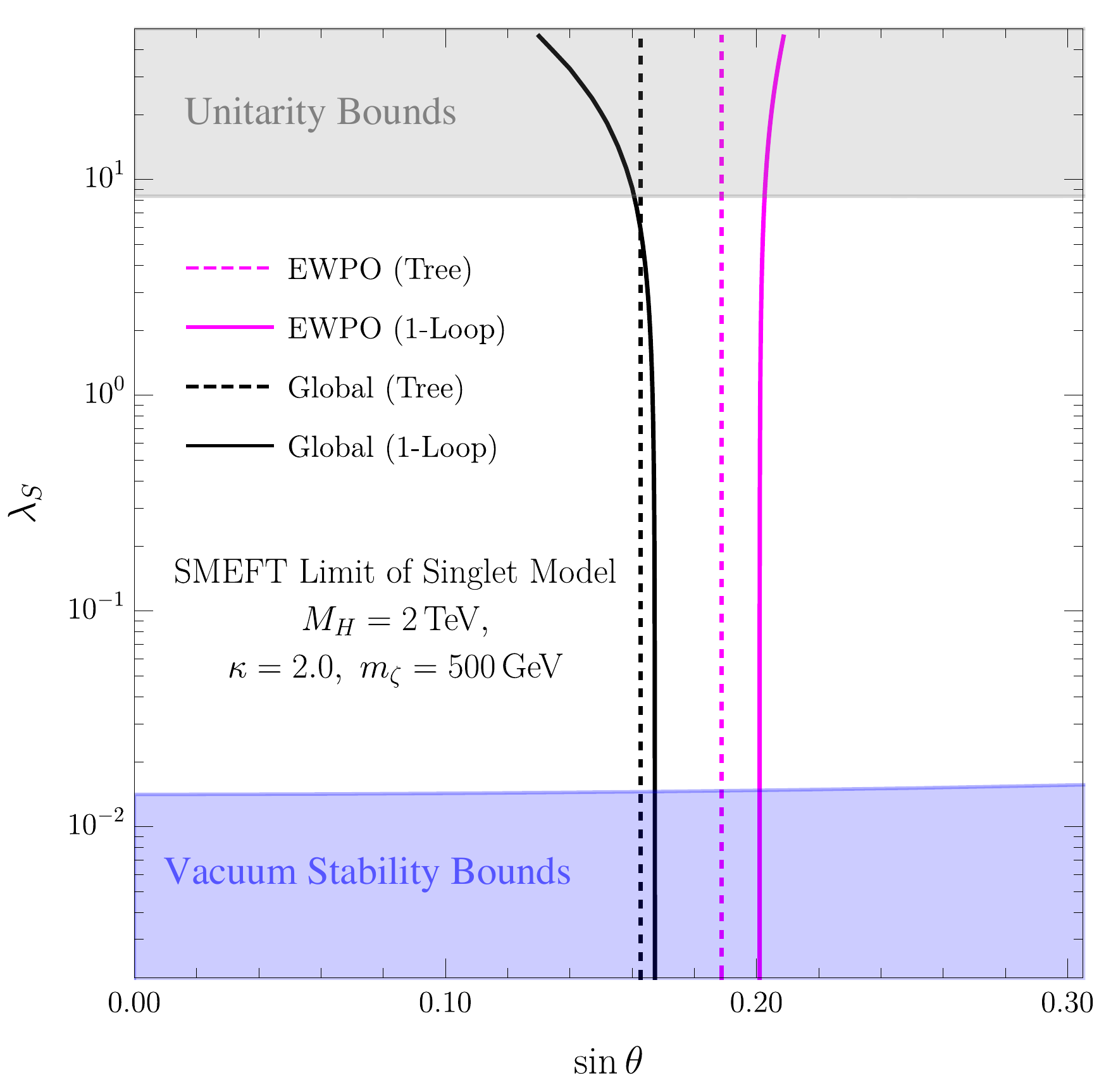}
\vskip -0.5cm
\caption{ 95\% C.L. limits on $\sin\theta$ as a function of $\kappa$ (LHS) and $\lambda_S$ (RHS) for fixed $M_H=2\,\textrm{TeV}$ and $m_{\zeta} = 500\,\textrm{GeV}$. The fits with tree-level matching are shown as dashed curves, with solid curves showing the 1-loop result.
The black curves show the result of a global fit to Higgs, diboson, and electroweak precision data, while the pink curves only the electroweak precision observables.
The regions to the right of the curves are excluded by the fits.
The grey and blue shaded regions are forbidden by unitarity and 
electroweak vacuum stability requirements, respectively.
}
\label{fig:globfits}
\vspace*{\floatsep}
\includegraphics[width=0.48\linewidth]{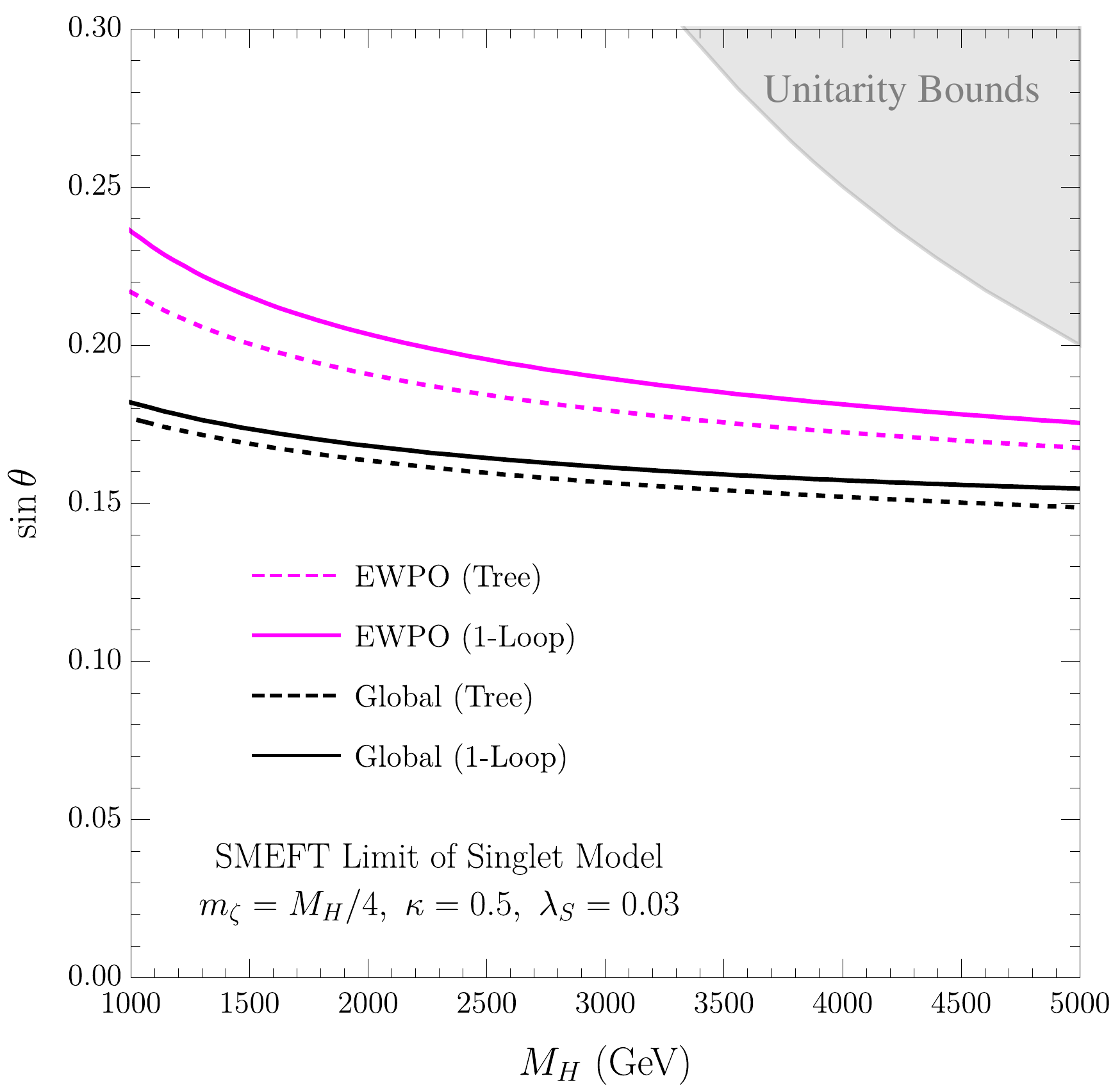}
~
\includegraphics[width=0.48\linewidth]{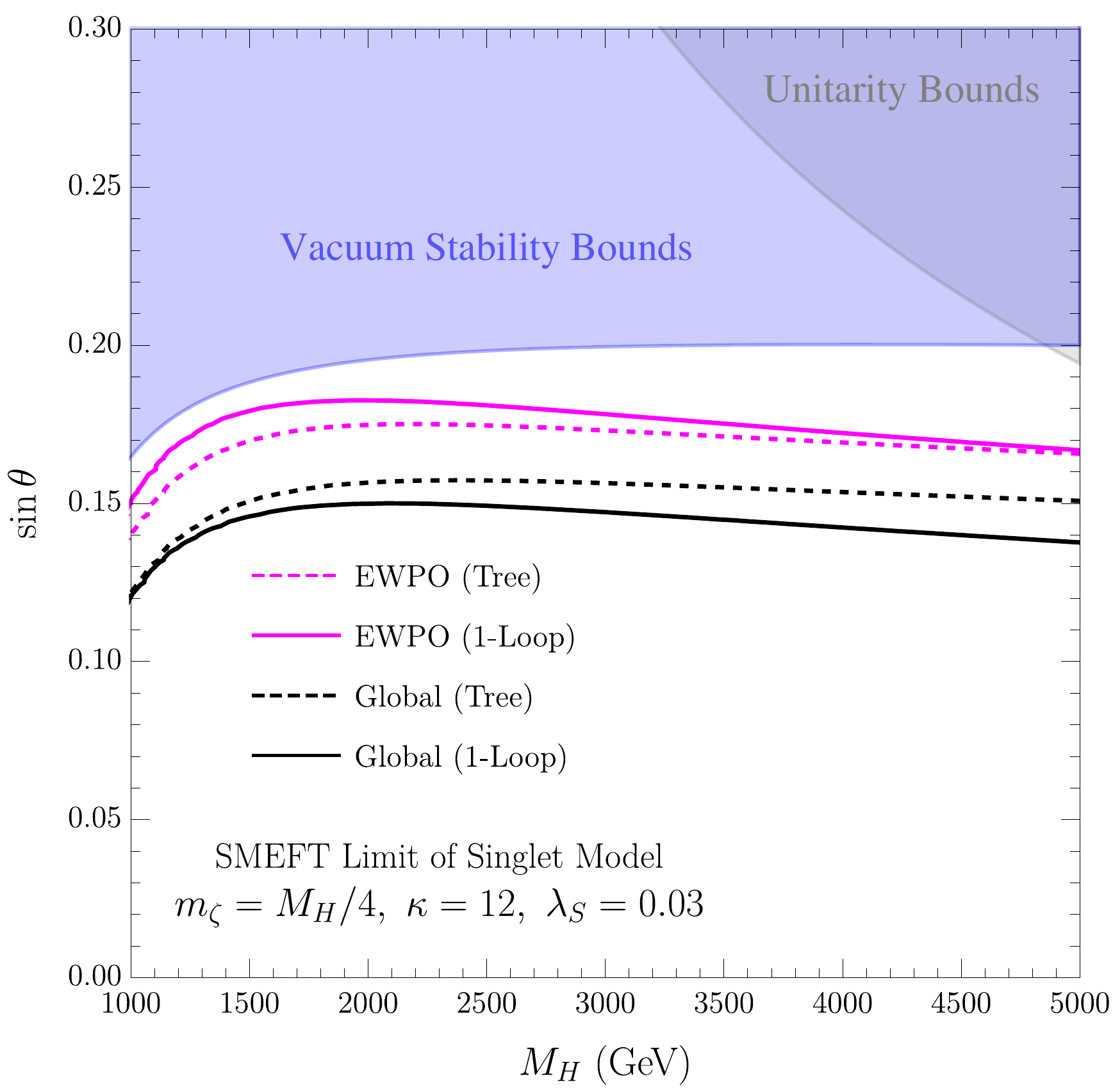}
\vskip -0.5cm
\caption{ As in Fig.~\ref{fig:globfits}, now showing limits on $\sin\theta$ as a function of the heavy Higgs mass, $M_H$, with fixed values of $\kappa$, $\lambda_S$ and $m_{\zeta} / M_H$.
Regions above the curves are excluded.
}
\label{fig:stmh}
\end{figure}

Figs.~\ref{fig:globfits} and \ref{fig:stmh} contain our major results.
In terms of the parameters of the singlet model given in Eq.\,\eqref{eq:pardef}, we fix $M_H=2\,\textrm{TeV}$ and determine the maximum allowed value of $\sin\theta$ in terms of the other unknown parameters of the model, $\lambda_S, ~\kappa,$ and $m_\zeta$.
The curves are relatively insensitive to $m_\zeta$ and $\lambda_S$ (RHS), and the major sensitivity is to $\kappa$ (LHS) of Fig.~\ref{fig:globfits}. 
We show the regions excluded by unitarity bounds and by vacuum stability bounds.
The black curves include the Higgs, diboson, and EWPO data.
For $\kappa \lesssim 8$, the inclusion of the 1-loop matching makes very little difference, but as $\kappa$ becomes large and approaches the unitarity bound, the difference between tree level and 1-loop matching can be of ${\cal{O}}(10\%)$.
We separately show the limits from only EWPO limits in magenta and note that the 1-loop matching slightly improves the bound on $\sin\theta$.

Another interesting way to look at the results is to look at the maximum allowed value of $\sin\theta$ as a function of the heavy Higgs mass, $M_H$, for fixed values of $\kappa,m_\zeta$ and $\lambda_S$ as shown in Fig.~\ref{fig:stmh}.
We see that including the 1-loop matching changes the bound on $\sin\theta$ only marginally. 
The effect is larger as $\kappa$ is increased.

Single parameter fits to models with an additional scalar have been presented in Ref.~\cite{Ellis:2018gqa} and updated in Ref.~\cite{Ellis:2020unq} using the dictionary of Ref.~\cite{deBlas:2017xtg}.
Assuming $C_{H\square}=-1/M^2$ and $C_H=0$, they find a limit $M_H> 900\,\textrm{GeV}$ at $2\sigma$.
Our tree level matching result of Ref.~\cite{Dawson:2020oco} is roughly compatible with this bound, although we find that the inclusion of the renormalization group running of the coefficients (in particular $C_{HD}$ which is generated by renormalization group running) is numerically significant, so the bounds cannot be directly compared.\footnote{Fits to the singlet model with 1-loop matching, but no renormalization group running, are given in  Ref.~\cite{Anisha:2020ggj}, but the results are not in a form that we can compare with.}

\section{Conclusions}
\label{sec:concl}

We have re-examined the sensitivity of a global fit to electroweak precision observables and to Higgs and diboson data on the parameters of a scalar singlet model in both the full UV complete model and in the low energy approximation where the heavy scalar is integrated out and the parameters are matched to the dimension-6 SMEFT.
In the full singlet model, we find equivalent limits on the allowed mixing angle from the complete EWPO fit and from the fit to the oblique parameters when $M_H$ is heavy.
For the case with the second Higgs boson much lighter than $M_Z$,
the oblique parameter limits are not a good approximation to the full fit.
When the new scalar is very heavy, we integrate it out and match to the dimension-6 SMEFT and then 
perform the global fit both using tree level and 1-loop matching at the high scale and derive limits on the parameters of the
singlet theory from the SMEFT fit.
We find that the effect on the fit of  including the 1-loop matching is never larger than ${\cal{O}}(10\%)$ and that the results are quite insensitive to variations in the singlet Lagrangian parameters other than the portal term, $\kappa$.

Digital data can be found at \url{https://quark.phy.bnl.gov/Digital_Data_Archive/dawson/singlet_21}.

\section*{Acknowledgements}

SD is supported by the United States Department of Energy under Grant Contract DE-SC0012704.
The work of PPG has received financial support from Xunta de Galicia (Centro singular de investigaci\'on de Galicia accreditation 2019-2022), by European Union ERDF, and by  ``Mar\'ia  de Maeztu"  Units  of  Excellence program  MDM-2016-0692  and  the Spanish Research State Agency. 
The work of SH was supported by DOE Grant DE-SC0013607 and by the Alfred P. Sloan Foundation Grant No. G-2019-12504.

\bibliographystyle{utphys}
\bibliography{singlet_global}

\providecommand{\href}[2]{#2}\begingroup\raggedright\begin{thebibliography}{10}

\bibitem{Bowen:2007ia}
M.~Bowen, Y.~Cui, and J.~D. Wells, ``{Narrow trans-TeV Higgs bosons and H
  $\rightarrow $ hh decays: Two LHC search paths for a hidden sector Higgs
  boson},'' \href{http://dx.doi.org/10.1088/1126-6708/2007/03/036}{{\em JHEP}
  {\bfseries 0703} (2007) 036},
\href{http://arxiv.org/abs/hep-ph/0701035}{{\ttfamily arXiv:hep-ph/0701035
  [hep-ph]}}.

\bibitem{O'Connell:2006wi}
D.~O'Connell, M.~J. Ramsey-Musolf, and M.~B. Wise, ``{Minimal Extension of the
  Standard Model Scalar Sector},''
  \href{http://dx.doi.org/10.1103/PhysRevD.75.037701}{{\em Phys. Rev.}
  {\bfseries D75} (2007) 037701},
\href{http://arxiv.org/abs/hep-ph/0611014}{{\ttfamily arXiv:hep-ph/0611014
  [hep-ph]}}.

\bibitem{Dawson:2015haa}
S.~Dawson and I.~M. Lewis, ``{NLO corrections to double Higgs boson production
  in the Higgs singlet model},''
  \href{http://dx.doi.org/10.1103/PhysRevD.92.094023}{{\em Phys. Rev.}
  {\bfseries D92} no.~9, (2015) 094023},
\href{http://arxiv.org/abs/1508.05397}{{\ttfamily arXiv:1508.05397 [hep-ph]}}.

\bibitem{Muhlleitner:2020wwk}
M.~M\"uhlleitner, M.~O. Sampaio, R.~Santos, and J.~Wittbrodt, ``{ScannerS:
  Parameter Scans in Extended Scalar Sectors},''
  \href{http://arxiv.org/abs/2007.02985}{{\ttfamily arXiv:2007.02985
  [hep-ph]}}.

\bibitem{Robens:2015gla}
T.~Robens and T.~Stefaniak, ``{Status of the Higgs Singlet Extension of the
  Standard Model after LHC Run 1},''
  \href{http://dx.doi.org/10.1140/epjc/s10052-015-3323-y}{{\em Eur.Phys.J.}
  {\bfseries C75} no.~3, (2015) 104},
\href{http://arxiv.org/abs/1501.02234}{{\ttfamily arXiv:1501.02234 [hep-ph]}}.

\bibitem{Costa:2015llh}
R.~Costa, M.~M\"uhlleitner, M.~O.~P. Sampaio, and R.~Santos, ``{Singlet
  Extensions of the Standard Model at LHC Run 2: Benchmarks and Comparison with
  the NMSSM},'' \href{http://dx.doi.org/10.1007/JHEP06(2016)034}{{\em JHEP}
  {\bfseries 06} (2016) 034}, \href{http://arxiv.org/abs/1512.05355}{{\ttfamily
  arXiv:1512.05355 [hep-ph]}}.

\bibitem{Chen:2014ask}
C.-Y. Chen, S.~Dawson, and I.~M. Lewis, ``{Exploring resonant di-Higgs boson
  production in the Higgs singlet model},''
  \href{http://dx.doi.org/10.1103/PhysRevD.91.035015}{{\em Phys. Rev.}
  {\bfseries D91} no.~3, (2015) 035015},
\href{http://arxiv.org/abs/1410.5488}{{\ttfamily arXiv:1410.5488 [hep-ph]}}.

\bibitem{Huber:2006wf}
S.~J. Huber, T.~Konstandin, T.~Prokopec, and M.~G. Schmidt, ``{Electroweak
  Phase Transition and Baryogenesis in the nMSSM},''
  \href{http://dx.doi.org/10.1016/j.nuclphysb.2006.09.003}{{\em Nucl. Phys. B}
  {\bfseries 757} (2006) 172--196},
  \href{http://arxiv.org/abs/hep-ph/0606298}{{\ttfamily arXiv:hep-ph/0606298}}.

\bibitem{Profumo:2007wc}
S.~Profumo, M.~J. Ramsey-Musolf, and G.~Shaughnessy, ``{Singlet Higgs
  phenomenology and the electroweak phase transition},''
  \href{http://dx.doi.org/10.1088/1126-6708/2007/08/010}{{\em JHEP} {\bfseries
  08} (2007) 010}, \href{http://arxiv.org/abs/0705.2425}{{\ttfamily
  arXiv:0705.2425 [hep-ph]}}.

\bibitem{Espinosa:2011ax}
J.~R. Espinosa, T.~Konstandin, and F.~Riva, ``{Strong Electroweak Phase
  Transitions in the Standard Model with a Singlet},''
  \href{http://dx.doi.org/10.1016/j.nuclphysb.2011.09.010}{{\em Nucl. Phys.}
  {\bfseries B854} (2012) 592--630},
\href{http://arxiv.org/abs/1107.5441}{{\ttfamily arXiv:1107.5441 [hep-ph]}}.

\bibitem{Barger:2011vm}
V.~Barger, D.~J.~H. Chung, A.~J. Long, and L.-T. Wang, ``{Strongly First Order
  Phase Transitions Near an Enhanced Discrete Symmetry Point},''
  \href{http://dx.doi.org/10.1016/j.physletb.2012.02.040}{{\em Phys. Lett. B}
  {\bfseries 710} (2012) 1--7},
  \href{http://arxiv.org/abs/1112.5460}{{\ttfamily arXiv:1112.5460 [hep-ph]}}.

\bibitem{Profumo:2014opa}
S.~Profumo, M.~J. Ramsey-Musolf, C.~L. Wainwright, and P.~Winslow,
  ``{Singlet-catalyzed electroweak phase transitions and precision Higgs boson
  studies},'' \href{http://dx.doi.org/10.1103/PhysRevD.91.035018}{{\em Phys.
  Rev.} {\bfseries D91} no.~3, (2015) 035018},
\href{http://arxiv.org/abs/1407.5342}{{\ttfamily arXiv:1407.5342 [hep-ph]}}.

\bibitem{Curtin:2014jma}
D.~Curtin, P.~Meade, and C.-T. Yu, ``{Testing Electroweak Baryogenesis with
  Future Colliders},'' \href{http://dx.doi.org/10.1007/JHEP11(2014)127}{{\em
  JHEP} {\bfseries 11} (2014) 127},
  \href{http://arxiv.org/abs/1409.0005}{{\ttfamily arXiv:1409.0005 [hep-ph]}}.

\bibitem{Kotwal:2016tex}
A.~V. Kotwal, M.~J. Ramsey-Musolf, J.~M. No, and P.~Winslow,
  ``{Singlet-catalyzed electroweak phase transitions in the 100 TeV
  frontier},'' \href{http://dx.doi.org/10.1103/PhysRevD.94.035022}{{\em Phys.
  Rev.} {\bfseries D94} no.~3, (2016) 035022},
\href{http://arxiv.org/abs/1605.06123}{{\ttfamily arXiv:1605.06123 [hep-ph]}}.

\bibitem{Huang:2017jws}
T.~Huang, J.~M. No, L.~Pernie, M.~Ramsey-Musolf, A.~Safonov, M.~Spannowsky, and
  P.~Winslow, ``{Resonant di-Higgs boson production in the $b{\bar b}WW$
  channel: Probing the electroweak phase transition at the LHC},''
  \href{http://dx.doi.org/10.1103/PhysRevD.96.035007}{{\em Phys. Rev.}
  {\bfseries D96} no.~3, (2017) 035007},
\href{http://arxiv.org/abs/1701.04442}{{\ttfamily arXiv:1701.04442 [hep-ph]}}.

\bibitem{Chen:2017qcz}
C.-Y. Chen, J.~Kozaczuk, and I.~M. Lewis, ``{Non-resonant Collider Signatures
  of a Singlet-Driven Electroweak Phase Transition},''
  \href{http://dx.doi.org/10.1007/JHEP08(2017)096}{{\em JHEP} {\bfseries 08}
  (2017) 096},
\href{http://arxiv.org/abs/1704.05844}{{\ttfamily arXiv:1704.05844 [hep-ph]}}.

\bibitem{Kurup:2017dzf}
G.~Kurup and M.~Perelstein, ``{Dynamics of Electroweak Phase Transition In
  Singlet-Scalar Extension of the Standard Model},''
  \href{http://dx.doi.org/10.1103/PhysRevD.96.015036}{{\em Phys. Rev. D}
  {\bfseries 96} no.~1, (2017) 015036},
  \href{http://arxiv.org/abs/1704.03381}{{\ttfamily arXiv:1704.03381
  [hep-ph]}}.

\bibitem{Li:2019tfd}
H.-L. Li, M.~Ramsey-Musolf, and S.~Willocq, ``{Probing a scalar
  singlet-catalyzed electroweak phase transition with resonant di-Higgs boson
  production in the $4b$ channel},''
  \href{http://dx.doi.org/10.1103/PhysRevD.100.075035}{{\em Phys. Rev.}
  {\bfseries D100} no.~7, (2019) 075035},
\href{http://arxiv.org/abs/1906.05289}{{\ttfamily arXiv:1906.05289 [hep-ph]}}.

\bibitem{Craig:2013xia}
N.~Craig, C.~Englert, and M.~McCullough, ``{New Probe of Naturalness},''
  \href{http://dx.doi.org/10.1103/PhysRevLett.111.121803}{{\em Phys. Rev.
  Lett.} {\bfseries 111} no.~12, (2013) 121803},
  \href{http://arxiv.org/abs/1305.5251}{{\ttfamily arXiv:1305.5251 [hep-ph]}}.

\bibitem{Curtin:2015bka}
D.~Curtin and P.~Saraswat, ``{Towards a No-Lose Theorem for Naturalness},''
  \href{http://dx.doi.org/10.1103/PhysRevD.93.055044}{{\em Phys. Rev. D}
  {\bfseries 93} no.~5, (2016) 055044},
  \href{http://arxiv.org/abs/1509.04284}{{\ttfamily arXiv:1509.04284
  [hep-ph]}}.

\bibitem{Silveira:1985rk}
V.~Silveira and A.~Zee, ``{Scalar Phantoms},''
  \href{http://dx.doi.org/10.1016/0370-2693(85)90624-0}{{\em Phys. Lett. B}
  {\bfseries 161} (1985) 136--140}.

\bibitem{McDonald:1993ex}
J.~McDonald, ``{Gauge singlet scalars as cold dark matter},''
  \href{http://dx.doi.org/10.1103/PhysRevD.50.3637}{{\em Phys. Rev. D}
  {\bfseries 50} (1994) 3637--3649},
  \href{http://arxiv.org/abs/hep-ph/0702143}{{\ttfamily arXiv:hep-ph/0702143}}.

\bibitem{Burgess:2000yq}
C.~P. Burgess, M.~Pospelov, and T.~ter Veldhuis, ``{The Minimal model of
  nonbaryonic dark matter: A Singlet scalar},''
  \href{http://dx.doi.org/10.1016/S0550-3213(01)00513-2}{{\em Nucl. Phys. B}
  {\bfseries 619} (2001) 709--728},
  \href{http://arxiv.org/abs/hep-ph/0011335}{{\ttfamily arXiv:hep-ph/0011335}}.

\bibitem{Menon:2004wv}
A.~Menon, D.~E. Morrissey, and C.~E.~M. Wagner, ``{Electroweak baryogenesis and
  dark matter in the nMSSM},''
  \href{http://dx.doi.org/10.1103/PhysRevD.70.035005}{{\em Phys. Rev. D}
  {\bfseries 70} (2004) 035005},
  \href{http://arxiv.org/abs/hep-ph/0404184}{{\ttfamily arXiv:hep-ph/0404184}}.

\bibitem{He:2008qm}
X.-G. He, T.~Li, X.-Q. Li, J.~Tandean, and H.-C. Tsai, ``{Constraints on Scalar
  Dark Matter from Direct Experimental Searches},''
  \href{http://dx.doi.org/10.1103/PhysRevD.79.023521}{{\em Phys. Rev. D}
  {\bfseries 79} (2009) 023521},
  \href{http://arxiv.org/abs/0811.0658}{{\ttfamily arXiv:0811.0658 [hep-ph]}}.

\bibitem{Gonderinger:2009jp}
M.~Gonderinger, Y.~Li, H.~Patel, and M.~J. Ramsey-Musolf, ``{Vacuum Stability,
  Perturbativity, and Scalar Singlet Dark Matter},''
  \href{http://dx.doi.org/10.1007/JHEP01(2010)053}{{\em JHEP} {\bfseries 01}
  (2010) 053}, \href{http://arxiv.org/abs/0910.3167}{{\ttfamily arXiv:0910.3167
  [hep-ph]}}.

\bibitem{Mambrini:2011ik}
Y.~Mambrini, ``{Higgs searches and singlet scalar dark matter: Combined
  constraints from XENON 100 and the LHC},''
  \href{http://dx.doi.org/10.1103/PhysRevD.84.115017}{{\em Phys. Rev. D}
  {\bfseries 84} (2011) 115017},
  \href{http://arxiv.org/abs/1108.0671}{{\ttfamily arXiv:1108.0671 [hep-ph]}}.

\bibitem{Dawson:2020oco}
S.~Dawson, S.~Homiller, and S.~D. Lane, ``{Putting standard model EFT fits to
  work},'' \href{http://dx.doi.org/10.1103/PhysRevD.102.055012}{{\em Phys. Rev.
  D} {\bfseries 102} no.~5, (2020) 055012},
  \href{http://arxiv.org/abs/2007.01296}{{\ttfamily arXiv:2007.01296
  [hep-ph]}}.

\bibitem{Brehmer:2015rna}
J.~Brehmer, A.~Freitas, D.~Lopez-Val, and T.~Plehn, ``{Pushing Higgs Effective
  Theory to its Limits},''
  \href{http://dx.doi.org/10.1103/PhysRevD.93.075014}{{\em Phys. Rev.}
  {\bfseries D93} no.~7, (2016) 075014},
\href{http://arxiv.org/abs/1510.03443}{{\ttfamily arXiv:1510.03443 [hep-ph]}}.

\bibitem{Henning:2014gca}
B.~Henning, X.~Lu, and H.~Murayama, ``{What do precision Higgs measurements buy
  us?},''
\href{http://arxiv.org/abs/1404.1058}{{\ttfamily arXiv:1404.1058 [hep-ph]}}.

\bibitem{Henning:2014wua}
B.~Henning, X.~Lu, and H.~Murayama, ``{How to use the Standard Model effective
  field theory},'' \href{http://dx.doi.org/10.1007/JHEP01(2016)023}{{\em JHEP}
  {\bfseries 01} (2016) 023}, \href{http://arxiv.org/abs/1412.1837}{{\ttfamily
  arXiv:1412.1837 [hep-ph]}}.

\bibitem{Gorbahn:2015gxa}
M.~Gorbahn, J.~M. No, and V.~Sanz, ``{Benchmarks for Higgs Effective Theory:
  Extended Higgs Sectors},''
  \href{http://dx.doi.org/10.1007/JHEP10(2015)036}{{\em JHEP} {\bfseries 10}
  (2015) 036},
\href{http://arxiv.org/abs/1502.07352}{{\ttfamily arXiv:1502.07352 [hep-ph]}}.

\bibitem{Ellis:2018gqa}
J.~Ellis, C.~W. Murphy, V.~Sanz, and T.~You, ``{Updated Global SMEFT Fit to
  Higgs, Diboson and Electroweak Data},''
  \href{http://dx.doi.org/10.1007/JHEP06(2018)146}{{\em JHEP} {\bfseries 06}
  (2018) 146}, \href{http://arxiv.org/abs/1803.03252}{{\ttfamily
  arXiv:1803.03252 [hep-ph]}}.

\bibitem{Ellis:2020unq}
J.~Ellis, M.~Madigan, K.~Mimasu, V.~Sanz, and T.~You, ``{Top, Higgs, Diboson
  and Electroweak Fit to the Standard Model Effective Field Theory},''
  \href{http://arxiv.org/abs/2012.02779}{{\ttfamily arXiv:2012.02779
  [hep-ph]}}.

\bibitem{Anisha:2020ggj}
Anisha, S.~Das~Bakshi, J.~Chakrabortty, and S.~K. Patra, ``{A Step Toward Model
  Comparison: Connecting Electroweak-Scale Observables to BSM through EFT and
  Bayesian Statistics},'' \href{http://arxiv.org/abs/2010.04088}{{\ttfamily
  arXiv:2010.04088 [hep-ph]}}.

\bibitem{Egana-Ugrinovic:2015vgy}
D.~Egana-Ugrinovic and S.~Thomas, ``{Effective Theory of Higgs Sector Vacuum
  States},'' \href{http://arxiv.org/abs/1512.00144}{{\ttfamily arXiv:1512.00144
  [hep-ph]}}.

\bibitem{Bakshi:2020eyg}
S.~Das~Bakshi, J.~Chakrabortty, and M.~Spannowsky, ``{Classifying Standard
  Model Extensions Effectively with Precision Observables},''
  \href{http://arxiv.org/abs/2012.03839}{{\ttfamily arXiv:2012.03839
  [hep-ph]}}.

\bibitem{Kribs:2017znd}
G.~D. Kribs, A.~Maier, H.~Rzehak, M.~Spannowsky, and P.~Waite, ``{Electroweak
  oblique parameters as a probe of the trilinear Higgs boson
  self-interaction},'' \href{http://dx.doi.org/10.1103/PhysRevD.95.093004}{{\em
  Phys. Rev. D} {\bfseries 95} no.~9, (2017) 093004},
  \href{http://arxiv.org/abs/1702.07678}{{\ttfamily arXiv:1702.07678
  [hep-ph]}}.

\bibitem{Falkowski:2015iwa}
A.~Falkowski, C.~Gross, and O.~Lebedev, ``{A second Higgs from the Higgs
  portal},'' \href{http://dx.doi.org/10.1007/JHEP05(2015)057}{{\em JHEP}
  {\bfseries 05} (2015) 057},
\href{http://arxiv.org/abs/1502.01361}{{\ttfamily arXiv:1502.01361 [hep-ph]}}.

\bibitem{Jiang:2018pbd}
M.~Jiang, N.~Craig, Y.-Y. Li, and D.~Sutherland, ``{Complete One-Loop Matching
  for a Singlet Scalar in the Standard Model EFT},''
  \href{http://dx.doi.org/10.1007/JHEP02(2019)031}{{\em JHEP} {\bfseries 02}
  (2019) 031}, \href{http://arxiv.org/abs/1811.08878}{{\ttfamily
  arXiv:1811.08878 [hep-ph]}}. [Erratum: JHEP 01, 135 (2021)].

\bibitem{Haisch:2020ahr}
U.~Haisch, M.~Ruhdorfer, E.~Salvioni, E.~Venturini, and A.~Weiler, ``{Singlet
  night in Feynman-ville: one-loop matching of a real scalar},''
  \href{http://dx.doi.org/10.1007/JHEP04(2020)164}{{\em JHEP} {\bfseries 04}
  (2020) 164}, \href{http://arxiv.org/abs/2003.05936}{{\ttfamily
  arXiv:2003.05936 [hep-ph]}}. [Erratum: JHEP 07, 066 (2020)].

\bibitem{Cohen:2020fcu}
T.~Cohen, X.~Lu, and Z.~Zhang, ``{Functional Prescription for EFT Matching},''
  \href{http://arxiv.org/abs/2011.02484}{{\ttfamily arXiv:2011.02484
  [hep-ph]}}.

\bibitem{Buchalla:2016bse}
G.~Buchalla, O.~Cata, A.~Celis, and C.~Krause, ``{Standard Model Extended by a
  Heavy Singlet: Linear vs. Nonlinear EFT},''
  \href{http://dx.doi.org/10.1016/j.nuclphysb.2017.02.006}{{\em Nucl. Phys.}
  {\bfseries B917} (2017) 209--233},
\href{http://arxiv.org/abs/1608.03564}{{\ttfamily arXiv:1608.03564 [hep-ph]}}.

\bibitem{Dawson:2015oha}
S.~Dawson, A.~Ismail, and I.~Low, ``{What\textquoteright{}s in the loop? The
  anatomy of double Higgs production},''
  \href{http://dx.doi.org/10.1103/PhysRevD.91.115008}{{\em Phys. Rev. D}
  {\bfseries 91} no.~11, (2015) 115008},
  \href{http://arxiv.org/abs/1504.05596}{{\ttfamily arXiv:1504.05596
  [hep-ph]}}.

\bibitem{Aad:2019mbh}
{\bfseries ATLAS} Collaboration, G.~Aad {\em et~al.}, ``{Combined measurements
  of Higgs boson production and decay using up to $80$ fb$^{-1}$ of
  proton-proton collision data at $\sqrt{s}=$ 13 TeV collected with the ATLAS
  experiment},'' \href{http://dx.doi.org/10.1103/PhysRevD.101.012002}{{\em
  Phys. Rev. D} {\bfseries 101} no.~1, (2020) 012002},
  \href{http://arxiv.org/abs/1909.02845}{{\ttfamily arXiv:1909.02845
  [hep-ex]}}.

\bibitem{CMS:2020gsy}
{\bfseries CMS} Collaboration, ``{Combined Higgs boson production and decay
  measurements with up to $137$ fb$^{-1}$ of proton-proton collision data at
  $\sqrt{s}$ = 13 TeV},''. \url{https://cds.cern.ch/record/2706103}.

\bibitem{Aad:2019uzh}
{\bfseries ATLAS} Collaboration, G.~Aad {\em et~al.}, ``{Combination of
  searches for Higgs boson pairs in $pp$ collisions at $\sqrt{s} = $13 TeV with
  the ATLAS detector},''
  \href{http://dx.doi.org/10.1016/j.physletb.2019.135103}{{\em Phys. Lett. B}
  {\bfseries 800} (2020) 135103},
  \href{http://arxiv.org/abs/1906.02025}{{\ttfamily arXiv:1906.02025
  [hep-ex]}}.

\bibitem{Hollik:1988ii}
W.~F.~L. Hollik, ``{Radiative Corrections in the Standard Model and their Role
  for Precision Tests of the Electroweak Theory},''
\href{http://dx.doi.org/10.1002/prop.2190380302}{{\em Fortsch. Phys.}
  {\bfseries 38} (1990) 165--260}.

\bibitem{Freitas:2014hra}
A.~Freitas, ``{Higher-order electroweak corrections to the partial widths and
  branching ratios of the Z boson},''
  \href{http://dx.doi.org/10.1007/JHEP04(2014)070}{{\em JHEP} {\bfseries 04}
  (2014) 070},
\href{http://arxiv.org/abs/1401.2447}{{\ttfamily arXiv:1401.2447 [hep-ph]}}.

\bibitem{Dawson:2019clf}
S.~Dawson and P.~P. Giardino, ``{Electroweak and QCD corrections to $Z$ and $W$
  pole observables in the standard model EFT},''
  \href{http://dx.doi.org/10.1103/PhysRevD.101.013001}{{\em Phys. Rev. D}
  {\bfseries 101} no.~1, (2020) 013001},
  \href{http://arxiv.org/abs/1909.02000}{{\ttfamily arXiv:1909.02000
  [hep-ph]}}.

\bibitem{Carrazza:2016gav}
S.~Carrazza, R.~K. Ellis, and G.~Zanderighi, ``{QCDLoop: a comprehensive
  framework for one-loop scalar integrals},''
  \href{http://dx.doi.org/10.1016/j.cpc.2016.07.033}{{\em Comput. Phys.
  Commun.} {\bfseries 209} (2016) 134--143},
\href{http://arxiv.org/abs/1605.03181}{{\ttfamily arXiv:1605.03181 [hep-ph]}}.

\bibitem{Lopez-Val:2014jva}
D.~Lopez-Val and T.~Robens, ``{$\Delta$ r and the W-boson mass in the singlet
  extension of the standard model},''
  \href{http://dx.doi.org/10.1103/PhysRevD.90.114018}{{\em Phys. Rev.}
  {\bfseries D90} no.~11, (2014) 114018},
\href{http://arxiv.org/abs/1406.1043}{{\ttfamily arXiv:1406.1043 [hep-ph]}}.

\bibitem{Dawson:2009yx}
S.~Dawson and W.~Yan, ``{Hiding the Higgs Boson with Multiple Scalars},''
  \href{http://dx.doi.org/10.1103/PhysRevD.79.095002}{{\em Phys. Rev.}
  {\bfseries D79} (2009) 095002},
\href{http://arxiv.org/abs/0904.2005}{{\ttfamily arXiv:0904.2005 [hep-ph]}}.

\bibitem{Englert:2020gcp}
C.~Englert, J.~Jaeckel, M.~Spannowsky, and P.~Stylianou, ``{Power meets
  Precision to explore the Symmetric Higgs Portal},''
  \href{http://dx.doi.org/10.1016/j.physletb.2020.135526}{{\em Phys. Lett. B}
  {\bfseries 806} (2020) 135526},
  \href{http://arxiv.org/abs/2002.07823}{{\ttfamily arXiv:2002.07823
  [hep-ph]}}.

\bibitem{Peskin:1991sw}
M.~E. Peskin and T.~Takeuchi, ``{Estimation of oblique electroweak
  corrections},'' \href{http://dx.doi.org/10.1103/PhysRevD.46.381}{{\em Phys.
  Rev. D} {\bfseries 46} (1992) 381--409}.

\bibitem{Zyla:2020zbs}
{\bfseries Particle Data Group} Collaboration, P.~Zyla {\em et~al.}, ``{Review
  of Particle Physics},'' \href{http://dx.doi.org/10.1093/ptep/ptaa104}{{\em
  PTEP} {\bfseries 2020} no.~8, (2020) 083C01}.

\bibitem{Ilnicka:2018def}
A.~Ilnicka, T.~Robens, and T.~Stefaniak, ``{Constraining Extended Scalar
  Sectors at the LHC and beyond},''
  \href{http://dx.doi.org/10.1142/S0217732318300070}{{\em Mod. Phys. Lett.}
  {\bfseries A33} no.~10n11, (2018) 1830007},
\href{http://arxiv.org/abs/1803.03594}{{\ttfamily arXiv:1803.03594 [hep-ph]}}.

\bibitem{Chalons:2016lyk}
G.~Chalons, D.~Lopez-Val, T.~Robens, and T.~Stefaniak, ``{The Higgs singlet
  extension at LHC Run 2},'' \href{http://dx.doi.org/10.22323/1.265.0113}{{\em
  PoS} {\bfseries DIS2016} (2016) 113},
\href{http://arxiv.org/abs/1606.07793}{{\ttfamily arXiv:1606.07793 [hep-ph]}}.

\bibitem{deBlas:2017wmn}
J.~de~Blas, M.~Ciuchini, E.~Franco, S.~Mishima, M.~Pierini, L.~Reina, and
  L.~Silvestrini, ``{The Global Electroweak and Higgs Fits in the LHC era},''
  \href{http://dx.doi.org/10.22323/1.314.0467}{{\em PoS} {\bfseries
  EPS-HEP2017} (2017) 467},
\href{http://arxiv.org/abs/1710.05402}{{\ttfamily arXiv:1710.05402 [hep-ph]}}.

\bibitem{Robens:2016xkb}
T.~Robens and T.~Stefaniak, ``{LHC Benchmark Scenarios for the Real Higgs
  Singlet Extension of the Standard Model},''
  \href{http://dx.doi.org/10.1140/epjc/s10052-016-4115-8}{{\em Eur. Phys. J. C}
  {\bfseries 76} no.~5, (2016) 268},
  \href{http://arxiv.org/abs/1601.07880}{{\ttfamily arXiv:1601.07880
  [hep-ph]}}.

\bibitem{Lee:1977yc}
B.~W. Lee, C.~Quigg, and H.~Thacker, ``{The Strength of Weak Interactions at
  Very High-Energies and the Higgs Boson Mass},''
  \href{http://dx.doi.org/10.1103/PhysRevLett.38.883}{{\em Phys. Rev. Lett.}
  {\bfseries 38} (1977) 883--885}.

\bibitem{Lee:1977eg}
B.~W. Lee, C.~Quigg, and H.~Thacker, ``{Weak Interactions at Very
  High-Energies: The Role of the Higgs Boson Mass},''
  \href{http://dx.doi.org/10.1103/PhysRevD.16.1519}{{\em Phys. Rev. D}
  {\bfseries 16} (1977) 1519}.

\bibitem{Dawson:2017vgm}
S.~Dawson and C.~W. Murphy, ``{Standard Model EFT and Extended Scalar
  Sectors},'' \href{http://dx.doi.org/10.1103/PhysRevD.96.015041}{{\em Phys.
  Rev.} {\bfseries D96} no.~1, (2017) 015041},
\href{http://arxiv.org/abs/1704.07851}{{\ttfamily arXiv:1704.07851 [hep-ph]}}.

\bibitem{Buchmuller:1985jz}
W.~Buchmuller and D.~Wyler, ``{Effective Lagrangian Analysis of New
  Interactions and Flavor Conservation},''
  \href{http://dx.doi.org/10.1016/0550-3213(86)90262-2}{{\em Nucl. Phys. B}
  {\bfseries 268} (1986) 621--653}.

\bibitem{Dedes:2017zog}
A.~Dedes, W.~Materkowska, M.~Paraskevas, J.~Rosiek, and K.~Suxho, ``{Feynman
  rules for the Standard Model Effective Field Theory in $R_{\xi}$-gauges},''
  \href{http://dx.doi.org/10.1007/JHEP06(2017)143}{{\em JHEP} {\bfseries 06}
  (2017) 143}, \href{http://arxiv.org/abs/1704.03888}{{\ttfamily
  arXiv:1704.03888 [hep-ph]}}.

\bibitem{deBlas:2017xtg}
J.~de~Blas, J.~C. Criado, M.~Perez-Victoria, and J.~Santiago, ``{Effective
  description of general extensions of the Standard Model: the complete
  tree-level dictionary},''
  \href{http://dx.doi.org/10.1007/JHEP03(2018)109}{{\em JHEP} {\bfseries 03}
  (2018) 109},
\href{http://arxiv.org/abs/1711.10391}{{\ttfamily arXiv:1711.10391 [hep-ph]}}.

\bibitem{Jenkins:2013zja}
E.~E. Jenkins, A.~V. Manohar, and M.~Trott, ``{Renormalization Group Evolution
  of the Standard Model Dimension Six Operators I: Formalism and lambda
  Dependence},'' \href{http://dx.doi.org/10.1007/JHEP10(2013)087}{{\em JHEP}
  {\bfseries 10} (2013) 087}, \href{http://arxiv.org/abs/1308.2627}{{\ttfamily
  arXiv:1308.2627 [hep-ph]}}.

\bibitem{Jenkins:2013wua}
E.~E. Jenkins, A.~V. Manohar, and M.~Trott, ``{Renormalization Group Evolution
  of the Standard Model Dimension Six Operators II: Yukawa Dependence},''
  \href{http://dx.doi.org/10.1007/JHEP01(2014)035}{{\em JHEP} {\bfseries 01}
  (2014) 035}, \href{http://arxiv.org/abs/1310.4838}{{\ttfamily arXiv:1310.4838
  [hep-ph]}}.

\bibitem{Alonso:2013hga}
R.~Alonso, E.~E. Jenkins, A.~V. Manohar, and M.~Trott, ``{Renormalization Group
  Evolution of the Standard Model Dimension Six Operators III: Gauge Coupling
  Dependence and Phenomenology},''
  \href{http://dx.doi.org/10.1007/JHEP04(2014)159}{{\em JHEP} {\bfseries 04}
  (2014) 159}, \href{http://arxiv.org/abs/1312.2014}{{\ttfamily arXiv:1312.2014
  [hep-ph]}}.

\bibitem{Degrassi:2017ucl}
G.~Degrassi, M.~Fedele, and P.~P. Giardino, ``{Constraints on the trilinear
  Higgs self coupling from precision observables},''
  \href{http://dx.doi.org/10.1007/JHEP04(2017)155}{{\em JHEP} {\bfseries 04}
  (2017) 155}, \href{http://arxiv.org/abs/1702.01737}{{\ttfamily
  arXiv:1702.01737 [hep-ph]}}.

\bibitem{Degrassi:2016wml}
G.~Degrassi, P.~P. Giardino, F.~Maltoni, and D.~Pagani, ``{Probing the Higgs
  self coupling via single Higgs production at the LHC},''
  \href{http://dx.doi.org/10.1007/JHEP12(2016)080}{{\em JHEP} {\bfseries 12}
  (2016) 080}, \href{http://arxiv.org/abs/1607.04251}{{\ttfamily
  arXiv:1607.04251 [hep-ph]}}.

\bibitem{Baglio:2019uty}
J.~Baglio, S.~Dawson, and S.~Homiller, ``{QCD corrections in Standard Model EFT
  fits to $WZ$ and $WW$ production},''
  \href{http://dx.doi.org/10.1103/PhysRevD.100.113010}{{\em Phys. Rev. D}
  {\bfseries 100} no.~11, (2019) 113010},
  \href{http://arxiv.org/abs/1909.11576}{{\ttfamily arXiv:1909.11576
  [hep-ph]}}.

\bibitem{Baglio:2020oqu}
J.~Baglio, S.~Dawson, S.~Homiller, S.~D. Lane, and I.~M. Lewis, ``{Validity of
  standard model EFT studies of VH and VV production at NLO},''
  \href{http://dx.doi.org/10.1103/PhysRevD.101.115004}{{\em Phys. Rev. D}
  {\bfseries 101} no.~11, (2020) 115004},
  \href{http://arxiv.org/abs/2003.07862}{{\ttfamily arXiv:2003.07862
  [hep-ph]}}.

\end{thebibliography}\endgroup

\end{document}